\documentclass[10pt, twocolumn, twoside]{IEEEtran}

\usepackage{cite}
\usepackage{enumerate}
\usepackage{graphicx}
\usepackage[cmex10]{amsmath} 
\usepackage{amssymb}
\usepackage{mathtools}
\usepackage{xspace}
\usepackage{subfigure}
\usepackage{multirow}
\usepackage[lined,linesnumbered,ruled,commentsnumbered]{algorithm2e} 
\usepackage{colonequals}
\usepackage[mathscr]{euscript}
\usepackage{amsthm}
\usepackage{mathrsfs}

\usepackage{epsfig}
\usepackage{epstopdf}

\usepackage{tikz}
\usepackage{pgfplots}

\usetikzlibrary{arrows,shapes,backgrounds,plotmarks,positioning}

\pgfplotsset{compat=newest}                         
\pgfplotsset{plot coordinates/math parser=false}
\newlength\figureheight
\newlength\figurewidth

\newtheorem{theorem}{Theorem}[section]

\newtheorem{lemma}[theorem]{Lemma}
\newtheorem{definition}[theorem]{Definition}
\newtheorem{corollary}[theorem]{Corollary}
\newtheorem{proposition}[theorem]{Proposition}

\newtheorem{example}[theorem]{Example}

\newcommand{\argmin}{\arg\!\min}

\newcommand{\op}{\text}
\newcommand{\RZ}[1]{\mathsf{Z}_{#1}}

\newcommand{\RW}[1]{\mathsf{W}_{#1}}

\newcommand{\RACO}{R_{\op{ACO}}}
\newcommand{\RNCO}{R_{\op{NCO}}}
\newcommand{\RRCO}{\mathscr{R}_{\op{CO}}}
\newcommand{\RRACO}{\mathscr{R}_{\op{ACO}}^*}
\newcommand{\RRNCO}{\mathscr{R}_{\op{NCO}}^*}
\newcommand{\SC}{\mathcal{C}_S}

\newcommand{\D}{\mathcal{L}}

\newcommand{\alphaComp}{\hat{\alpha}}

\newcommand{\SCal}{\mathcal{S}}

\newcommand{\HPat}[1]{\hat{\mathcal{P}}_{#1}}

\newcommand{\CoordSatCapFus}{\text{CoordSatCap}}
\newcommand{\StrMap}{\text{StrMap}}
\newcommand{\StrMapKomo}{\text{StrMapKolmogorov}}
\newcommand{\StrMapDistPAR}{\text{StrMapDistPAR}}
\newcommand{\PAR}{\text{PAR}}
\newcommand{\DistrPAR}{\text{DistrPAR}}
\newcommand{\MaxFlow}{\text{MaxFlow}}
\newcommand{\DA}{\text{DA}}

\newcommand{\Set}[1]{\{#1\}}

\newcommand{\ASet}[2]{\langle #1 \rangle_{#2}}

\newcommand{\Pat}{\mathcal{P}}
\newcommand{\Qat}[2]{\mathcal{Q}_{#1,#2}}
\newcommand{\X}{\mathcal{X}}
\newcommand{\TX}{\tilde{\mathcal{X}}}
\newcommand{\Y}{\mathcal{Y}}
\newcommand{\TY}{\tilde{\mathcal{Y}}}
\newcommand{\M}{\mathcal{M}}
\newcommand{\TM}{\tilde{\mathcal{M}}}
\newcommand{\N}{\mathcal{N}}
\newcommand{\TN}{\tilde{\mathcal{N}}}

\newcommand{\U}[2]{\mathcal{U}_{#1,#2}}
\newcommand{\TU}[2]{\tilde{\mathcal{U}}_{#1,#2}}

\newcommand{\Up}[1]{\mathcal{S}_{#1}}
\newcommand{\TUp}[1]{\tilde{{\mathcal{S}}}_{#1}}
\newcommand{\SU}{\mathcal{S}}
\newcommand{\TSU}{\tilde{\mathcal{S}}}
\newcommand{\UKomop}[1]{S_{#1}} 

\newcommand{\Patp}[1]{\mathcal{P}^{(#1)}}
\newcommand{\alphap}[1]{\alpha^{(#1)}}
\newcommand{\alphaU}{\underline{\alpha}}

\newcommand{\lambdap}[1]{\lambda^{(#1)}}

\newcommand{\rv}{\mathbf{r}} 
\newcommand{\wv}{\mathbf{w}} 

\newcommand{\Fu}[1]{f_{#1}}
\newcommand{\FuHat}[1]{\hat{f}_{#1}}
\newcommand{\FuU}[1]{g_{#1}}
\newcommand{\Cut}{\kappa} 
\newcommand{\FuUK}[1]{h_{#1}} 

\newcommand{\Real}{\mathbb{R}}
\newcommand{\RealP}{\mathbb{R}_{+}}    
\newcommand{\RealPP}{\mathbb{R}_{++}}    
\newcommand{\Z}{\mathbb{Z}}            

\newcommand{\SFM}{\text{SFM}}

\begin{document}

\title{Part I: Improving Computational Efficiency of Communication for Omniscience}

%

\author{Ni~Ding,~\IEEEmembership{Member,~IEEE}, Parastoo~Sadeghi,~\IEEEmembership{Senior Member,~IEEE}, and Thierry~Rakotoarivelo,~\IEEEmembership{Member,~IEEE}

\thanks{The primary results of this paper have been partly published in \cite{Ding2018Allerton}.}
\thanks{Ni Ding and Thierry Rakotoarivelo (email: $\{$ni.ding, thierry.rakotoarivelo$\}$@data61.csiro.au) are with Data61, 5/13 Garden Street, Eveleigh, NSW 2015.}
\thanks{Parastoo Sadeghi (email: $\{$parastoo.sadeghi$\}$@anu.edu.au) is with Research School of Electrical, Energy and Materials Engineering (RSEEME), The Australian National University, ACT, 2601. }
}

\markboth{IEEE Transactions}%
{Ding \MakeLowercase{\emph{et al.}}: Improving Computational Efficiency of Communication for Omniscience}

\maketitle

\begin{abstract}
    Communication for omniscience (CO) refers to the problem where the users in a finite set $V$ observe a discrete multiple random source and want to exchange data over broadcast channels to reach omniscience, the state where everyone recovers the entire source.
    This paper studies how to improve the computational complexity for the problem of minimizing the sum-rate for attaining omniscience in $V$. While the existing algorithms rely on the submodular function minimization (SFM) techniques and complete in $O(|V|^2 \cdot \SFM(|V|)$ time, we prove the strict strong map property of the nesting SFM problem. We propose a parametric ($\PAR$) algorithm that utilizes the parametric SFM techniques and reduce the the complexity to $O(|V| \cdot \SFM(|V|)$.
    The output of the $\PAR$ algorithm is in fact the segmented Dilworth truncation of the residual entropy for all minimum sum-rate estimates $\alpha$, which characterizes the principal sequence of partitions (PSP) and  solves some related problems: It not only determines the secret capacity, a dual problem to CO, and the network strength of a graph, but also outlines the hierarchical solution to a combinatorial clustering problem.
\end{abstract}


\section{introduction}

Let there be a finite number of users indexed by the set $V$. Each user observes a distinct component of a discrete memoryless multiple random source in private. The users are allowed to exchange their observations over public noiseless broadcast channels so as to attain \emph{omniscience}, the state that each user reconstructs all components in the multiple source. This process is called \emph{communication for omniscience} (CO) \cite{Csiszar2004}, where the fundamental problem is how to attain omniscience with the minimum sum of broadcast rates.
While the CO problem formulated in \cite{Csiszar2004} considers the asymptotic limits as the observation length goes to infinity, a non-asymptotic model is studied in \cite{Niti2010,Chan2011ITW,ChanSuccessiveIT}, in which, the number of observations is finite and the communication rates are restricted to be integral. The CO problem has a wide range of important applications, special cases, extensions, duals and interpretations.

The CO problem is dual with the secret capacity~\cite{Csiszar2004}, which is the maximum amount of secret key that can be generated by the users in $V$ and equals to the amount of information in the entire source, $H(V)$, subtracted by the minimum sum-rate in CO. A special case of CO is called coded cooperative data exchange (CCDE) \cite{Roua2010,Court2010,Court2010M,SprintRand2010,Ozgul2011,AbediniNonMinRank2012,Court2011,CourtIT2014},  where a group of users obtain parts of a packet set, say, via base-to-peer (B2P) transmissions. By broadcasting linear combinations of packets over peer-to-peer (P2P) channels, they help each other recover the entire packet set based on a suitable network coding scheme, e.g.,  the random linear network coding~\cite{SprintRand2010}.
It is shown in \cite{CourtIT2014,MiloIT2016,ChanMMI,Ding2016NetCod,Ding2018IT} that the solutions to the secret key agreement problem, CO and CCDE rely on the submodular function minimization (SFM) techniques in combinatorial optimization~\cite{Fujishige2005}. In a nutshell, all solutions in \cite{CourtIT2014,MiloIT2016,ChanMMI,Ding2016NetCod,Ding2018IT} come down to $O(|V |^2)$ calls of solving the SFM problem. Since the polynomial order of solving the SFM is still considerably high \cite[Chapter~VI]{Fujishige2005}, it is important to study whether the order-wise complexity $|V|^2$ in the computational complexity can be further reduced. This requires a deep understanding of the structure of the CO problem and its optimal solution. It is known from previous works \cite{ChanMMI,Ding2018IT} that the first critical/turning point in the principal sequence of partitions (PSP), a partition chain that is induced by the Dilworth truncation of the residual entropy function, plays a central role in solving the CO problem. This is essentially the first or coarsest partition in the PSP that is strictly finer than the partition $\Set{V}$.

Another important interpretation of CO is in the extension of the Shannon's mutual information to the multivariate case and is called the multivariate mutual information $I(V)$ \cite{ChanMMI}: $I(V)$ equals to the secret capacity.
This measure was used in~\cite{Chan2016InfoClustering} to interpret the PSP as a hierarchical clustering result: the partitions in the PSP contain the largest user subsets $X$ with $I (X )$ strictly greater than a given similarity threshold and get coarser (from bottom to top) as this similarity threshold decreases. This coincides with a more general combinatorial clustering framework, the minimum average clustering (MAC) in \cite{MinAveCost}, where both the entropy and cut functions are viewed as the inhomogeneity measure of a dataset. For the cut function, the first critical value in the PSP identifies the the network strength \cite{MinAveCost,Cunningham1985NetStrength} and the maximum number of edge-disjoint spanning trees \cite{IBMRep2011}. This, in return, well explains why the secret agreement problem in the pairwise independent network (PIN) source model, which has a graphical representation, can be solved by the tree packing algorithms in \cite{GSK2007PIN,PIN2010,PINSteinerTree2010}. Thus, instead of only focusing on one critical point for solving the minimum sum rate problem, it is also worth studying how to improve the existing complexity $O(|V |^2 \cdot \SFM(|V |))$ for determining the whole PSP.


\subsection{Contributions}

In this paper, we propose a parametric ($\PAR$) algorithm that reduces the complexity for solving the minimum sum-rate problem in CO and determining the PSP to $O(|V| \cdot \SFM(|V|))$.
The study starts with a review of the coordinate saturation capacity ($\CoordSatCapFus$) algorithm in \cite[Algorithm~3]{Ding2018IT}, which is a nesting algorithm in the modified decomposition algorithm (MDA) algorithm \cite[Algorithm~1]{Ding2018IT} that determines the Dilworth truncation for a given minimum sum-rate estimate $\alpha$.
We prove that the SFM problem in each iteration of $\CoordSatCapFus$ exhibits the strict strong map property in $\alpha$, based on which, a $\StrMap$ algorithm is proposed that determines the minimizer of this SFM problem for all values of $\alpha$.
The $\StrMap$ can be implemented by the existing parametric SFM (PSFM) algorithms \cite{Fleischer2003PSFM,Nagano2007PSFM,IwataPSFM1997} that complete at the same time as the SFM algorithm.

Based on the idea of $\CoordSatCapFus$, we propose a $\PAR$ algorithm that iteratively calls the subroutine $\StrMap$ to update the segmented minimizer of the Dilworth truncation for all values of the minimum sum-rate estimate $\alpha$. The critical/turning points of $\alpha$ as well as the corresponding minimizers/partitions, which characterize the segmented Dilworth truncation, converge to the PSP of $V$, where the first critical value determines the minimum sum-rate for both asymptotic and non-asymptotic model.
The $\PAR$ algorithm also outputs a segmented, or piecewise linear, rate vector $\rv_{\alpha,V} = (r_{\alpha,i} \colon i \in V)$ in $\alpha$ that determines an optimal rate vector for both asymptotic and non-asymptotic source models.
In addition, by choosing a proper order of iterations in the $\PAR$ algorithm, the optimal rate vector also minimizes the weighed minimum sum-rate in the optimal rate vector set.

The $\PAR$ algorithm invokes $|V|$  calls of $\StrMap$ and its complexity is $O(|V| \cdot \SFM(|V|))$.
It also allows distributed computation and can be applied to submodular functions other than the entropy function, e.g., the cut function.
The returned PSP solves the information-theoretic and MAC clustering problems in \cite{Chan2016InfoClustering} and \cite{MinAveCost}, respectively, with the complexity reduced from the existing algorithms by a factor of $|V|$.
For the cut function, the first critical point of the returned PSP determines the network strength, also the value of the secret capacity in the PIN model.
The work also studies another parametric algorithm proposed in \cite[Fig.~3]{Kolmogorov2010} for determining the PSP of the graph model. It is revealed that \cite[Fig.~3]{Kolmogorov2010} utilizes a non-strict strong map property, based on which, we propose a $\StrMap$ subroutine specifically for \cite[Fig.~3]{Kolmogorov2010} so that it also applies to any submodular function other than the cut function in a graph.

The proposed PAR algorithm also solves a successive omniscience (SO) problem, where the omniscience process takes stages: a user subset attains the local omniscience each time.
In Part II \cite{DingITSO2019} of this paper, we derive the achievability conditions for the multi-stage SO for both asymptotic and non-asymptotic models. By using the segmented Dilworth truncation and rate vector $\rv_{\alpha,V}$ returned by PAR, we propose algorithms extracting the user subset and achievable rate vector for each stage of SO such that, at the final stage, the global omniscience in $V$ is attained by the minimum sum-rate.
%

\subsection{Organization}

The rest of paper is organized as follows. The system model for CO is described in Section~\ref{sec:system}, where we also introduce the notation, review the existing results for the minimum sum-rate problem, including the PSP, and derive the properties of the $\CoordSatCapFus$ algorithm.
In Section~\ref{sec:ParAlgo}, we prove the strict strong map property and propose the $\PAR$ algorithm and its subroutine $\StrMap$ algorithm.
In Section~\ref{sec:Relation}, we discuss how the $\PAR$ algorithm contributes to the secret agreement, network attack and combinatorial clustering problems, where the relationship between the network strength and secret capacity in the PIN model is also explained.
In Section~\ref{sec:Distr}, we propose a distributed computation method of the $\PAR$ algorithm.

\section{System Model}
\label{sec:system}

Let $V$ with $|V|>1$ be a finite set that contains all users in the system. We call $V$ the \emph{ground set}. Let $\RZ{V}=(\RZ{i}:i\in V)$ be a vector of discrete random variables indexed by $V$. For each $i\in V$, user $i$ privately observes an $n$-sequence $\RZ{i}^n$ of the random source $\RZ{i}$ that is i.i.d.\ generated according to the joint distribution $P_{\RZ{V}}$. We allow users to exchange their observed data directly to recover the source sequence $\RZ{V}^n$. The state that each user obtains the total information in the entire multiple source is called \emph{omniscience}, and the process that users communicate with each other to attain omniscience is called \emph{communication for omniscience} (CO) \cite{Csiszar2004}.

Let $\rv_V=(r_i:i\in V)$ be a rate vector indexed by $V$. We call $\rv_V$ an \emph{achievable rate vector} if the omniscience can be attained by letting users communicate at the rates designated by $\rv_V$.
For the original CO problem formulated in \cite{Csiszar2004} considering the asymptotic limits as the \emph{block length} $n$ goes to infinity, each dimension $r_i$ is the compression rate denoting the expected code length at which user $i$ encode his/her observations. 
We also study a \emph{non-asymptotic model}, where $n$ is assumed to be finite. The \emph{finite linear source model} \cite{Chan2011ITW} is one of the non-asymptotic models, in which the multiple random source is represented by a vector that belongs to a finite field and each $r_i$ denotes the integer number of linear combinations of observations transmitted by user $i$.
This finite linear source model is of particular interest in that it models the CCDE problem \cite{Roua2010,Court2010,Court2010M} where the users communicate over P2P channels to help each other recover a packet set.
In this paper, for the omniscience problem in the non-asymptotic model, we focus on the finite linear source model. Therefore, we use the term non-asymptotic model, finite linear source model and CCDE interchangeably.

\subsection{Minimum Sum-rate Problem}
\label{subsec:MinSumRate}

For a given rate vector $\rv_V$, let $r \colon 2^V \mapsto \RealP$ be the \emph{sum-rate function} such that
$$ r(X)=\sum_{i\in X} r_i, \quad \forall X \subseteq V $$
with the convention $r(\emptyset)=0$.
The \emph{achievable rate region} is characterized in \cite{Csiszar2004} by the set of multiterminal Slepian-Wolf constraints \cite{SW1973,Cover1975}:
$$ \RRCO(V)=\Set{ \rv_V\in\Real^{|V|} \colon r(X) \geq H(X|V\setminus X),\forall X \subsetneq V },$$
where $H(X)$ is the amount of randomness in $\RZ{X}$ measured by the Shannon entropy \cite{Cover2012ITBook} and $H(X|Y)=H(X \cup Y)-H(Y)$ is the conditional entropy of $\RZ{X}$ given $\RZ{Y}$. In a finite linear source model, the entropy function $H$ reduces to the rank of a matrix that only takes integral values.

The fundamental problem in CO is to minimize the sum-rate in the achievable rate region \cite[Proposition 1]{Csiszar2004}
\begin{subequations}\label{eq:MinSumRate}
    \begin{align}
        \RACO(V) & = \min\Set{ r(V) \colon \rv_V \in \RRCO(V)}, \label{eq:MinSumRateACO}\\
        \RNCO(V) & = \min\Set{ r(V) \colon \rv_V \in \RRCO(V) \cap \Z^{|V|}},\label{eq:MinSumRateNCO}
    \end{align}
\end{subequations}
for the asymptotic and non-asymptotic models, respectively.
Denote by $\RRACO(V) = \Set{\rv_V \in \Real^{|V|} \colon r(V) = \RACO(V)}$ and $\RRNCO(V) = \Set{\rv_V \in \Z^{|V|} \colon r(V) = \RNCO(V)}$ the \emph{optimal rate vector set} for the asymptotic and non-asymptotic models, respectively.
We say that the minimum sum-rate problem is solved if the value of the minimum sum-rate in \eqref{eq:MinSumRate}, as well as an optimal rate vector are determined.

To efficiently solve the \emph{minimum sum-rate problem} without dealing with the exponentially growing number of constraints in the linear programming, \eqref{eq:MinSumRateACO} and \eqref{eq:MinSumRateNCO} are respectively converted to~\cite[Example 4]{Csiszar2004} \cite{Chan2008tight} \cite[Corollary 6]{Ding2018IT}
\begin{subequations} \label{eq:MinSumRatePat}
        \begin{align}
            \RACO(V) & = \max_{\Pat \in \Pi(V) \colon |\Pat| > 1} \sum_{C \in \Pat} \frac{H(V) - H(C)}{|\Pat|-1}, \label{eq:MinSumRateACOPat} \\
            \RNCO(V) & = \Big\lceil \max_{\Pat \in \Pi(V) \colon |\Pat| > 1} \sum_{C \in \Pat} \frac{H(V) - H(C)}{|\Pat|-1} \Big\rceil, \label{eq:MinSumRateNCOPat}
        \end{align}
\end{subequations}
where $\Pi(V)$ denotes the set containing all partitions of $V$. It is shown in \cite{ChanMMI,Ding2016NetCod,Ding2018IT} that the combinatorial optimization problem in \eqref{eq:MinSumRatePat} can be solved based on the existing submodular function minimization (SFM) techniques in polynomial time $O(|V|^2 \cdot \SFM(|V|))$.

\subsection{Existing Results}
\label{subsec:ExResults}

The efficiency for solving the minimum sum-rate problems in \eqref{eq:MinSumRatePat} relies on the submodularity of the entropy function $H$ and the induced structure in the partition lattice.
It is shown in \cite{Ding2018IT} that the validity of the algorithms proposed in \cite[Appendix~F]{CourtIT2014} and \cite[Algorithm~3]{MiloIT2016} for solving \eqref{eq:MinSumRateNCOPat} in CCDE and the MDA algorithm proposed in \cite[Algorithm ~1]{Ding2018IT} for solving both \eqref{eq:MinSumRateACOPat} and \eqref{eq:MinSumRateNCOPat} can be explained by the Dilworth truncation and the partition chain it forms in the estimation of $\RACO(V)$ or $\RNCO(V)$, which is called the principal sequence of partitions (PSP).
In this section, we introduce the notation and review the Dilworth truncation, PSP and the coordinate-wise saturation capacity ($\CoordSatCapFus$) algorithm, an essential nesting algorithm in \cite[Appendix~F]{CourtIT2014}, \cite[Algorithm~3]{MiloIT2016} and \cite[Algorithm ~1]{Ding2018IT}. The purpose is to summarize the existing results that are required to prove the strict strong map property in Section~\ref{sec:ParAlgo}.

\subsubsection{Preliminaries}

For $X \subseteq V$, let $\chi_X = (e_i \colon i \in V )$ be the \emph{characteristic vector} of the subset $X$ such that $e_i = 1$ if $i \in X$ and $e_i = 0$ if $i \notin X$. The notation $\chi_{\Set{i}}$ is simplified by $\chi_i$.
Let $\sqcup$ denote the disjoint union. For $\X$ that contains disjoint subsets of $V$, we denote by $\TX = \sqcup_{C \in \X} C$ the \emph{fusion} of $\X$. For example, for $\X = \Set{\Set{3,4},\Set{2},\Set{8}}$, $\TX = \Set{2,3,4,8}$.

For partitions $\Pat,\Pat' \in \Pi(V)$, we denote by $\Pat \preceq \Pat'$ if $\Pat$ is finer than $\Pat'$ and $\Pat \prec \Pat'$ if $\Pat$ is strictly finer than $\Pat'$.\footnote{The partition $\Pat$ is finer than $\Pat'$, if each subset in $\Pat$ is contained in some subset in $\Pat'$. }
For any $X \subseteq V$ and $\Pat \in \Pi(V)$, $\ASet{X}{\Pat} = \Set{X \cap C \colon C \in \Pat}$ denotes the decomposition of $X$ by $\Pat$. For example, for $X = \Set{1,2,4}$ and $\Pat = \Set{\Set{1,2,3},\Set{4}}$, $\ASet{X}{\Pat} = \Set{\Set{1,2},\Set{4}}$.

A function $f \colon 2^V \mapsto \Real$ is \emph{submodular} if $f(X) + f(Y) \geq f(X \cap Y) + f(X \cup Y)$ for all $X,Y \subseteq V$. The problem $\min \Set{f(X) \colon X \subseteq V}$ is a submodular function minimization (SFM) problem. It can be solved in strongly polynomial time and the set of minimizers $\argmin \Set{f(X) \colon X \subseteq V}$ form a set lattice such that the smallest/minimal minimizer $\bigcap \argmin \Set{f(X) \colon X \subseteq V}$ and largest/maximal minimizer $\bigcup \argmin \Set{f(X) \colon X \subseteq V}$ uniquely exist and can be determined at the same time when the SFM problem is solved \cite[Chapter~VI]{Fujishige2005}.

We call $\Phi = (\phi_1,\dotsc,\phi_{|V|})$ a \emph{linear ordering/permutation} of the indices in $V$ if $\phi_i \in V$ and $\phi_i \neq \phi_{i'}$ for all $i,i'\in\Set{1,\dotsc,|V|}$ such that $i\neq i'$.
For $i \in V$, let $V_i = \Set{\phi_1,\dotsc,\phi_i}$ be the set of the first $i$ users in the linear ordering $\Phi$.
We call $f^{V_i} \colon 2^{V_i} \mapsto \Real$ such that $f^{V_i}(X) = f(X)$ for all $X \subseteq V_i$ the \emph{reduction} of $f$ on $V_i$ \cite[Section~3.1(a)]{Fujishige2005}.
For example, for $\Phi = (2,3,1,4)$, $V_2 = \Set{2,3}$ and the reduction of $f$ on $V_2$ is $f^{V_2}(X) = f(X)$ for all $X \subseteq \Set{2,3}$.

\subsubsection{Dilworth Truncation}

Let $\alpha \in \RealP$ be an estimation of the minimum sum-rate and define a set function $\Fu{\alpha} \colon 2^{V} \mapsto \Real$ such that $\Fu{\alpha}(X) = \alpha - H(V) +H(X), \forall X \subseteq V$ except that $f(\emptyset) = 0$. This function is the same as the residual entropy function in \cite{ChanMMI} in that it offsets/subtracts the information amount in each nonempty subset $X$ by $H(V) - \alpha$.
Let $\Fu{\alpha}[\cdot]$ be a partition function such that $\Fu{\alpha}[\Pat] = \sum_{C \in \Pat} \Fu{\alpha}(C)$ for all $\Pat \in \Pi(V)$. The Dilworth truncation of $\Fu{\alpha}$ is \cite{Dilworth1944}
\begin{equation} \label{eq:Dilworth}
    \FuHat{\alpha}(V) =  \min_{\Pat \in \Pi(V)} \Fu{\alpha}[\Pat].
\end{equation}
The solution to \eqref{eq:Dilworth} exhibits a strong structure in $\alpha$ that is characterized by the PSP.

\subsubsection{Principal Sequence of Partitions (PSP)}
\label{subsec:PSP}

For a given $\alpha$, let $\Qat{\alpha}{V} = \bigwedge \argmin_{\Pat \in \Pi(V)} \Fu{\alpha}[\Pat]$ be the finest minimizer of \eqref{eq:Dilworth}.\footnote{The minimizers of \eqref{eq:Dilworth} form a partition lattice such that the finest and coarsest minimizers uniquely exist \cite{Narayanan1991PLP}.}
The value of Dilworth truncation $\FuHat{\alpha}(V)$ is piecewise linear strictly increasing in $\alpha$. It is determined by $p < |V| $ critical points
\begin{equation} \label{eq:PSPalpha}
    0 \leq \alphap{p} < \dotsc < \alphap{1} < \alphap{0} = H(V)
\end{equation}
with the corresponding finest minimizer $\Patp{j} = \Qat{\alphap{j}}{V} = \bigwedge \argmin_{\Pat \in \Pi(V)} \Fu{\alphap{j}}[\Pat]$ for all $j \in \Set{0,\dotsc,p}$ forming a partition chain
\begin{equation} \label{eq:PSPPat}
    \Set{\Set{i} \colon i\in V} = \Patp{p} \prec \dotsc \prec \Patp{1} \prec \Patp{0} = \Set{V}
\end{equation}
such that $\Qat{\alpha}{V} = \Patp{p}$ for $\alpha \in [0, \alphap{p}]$ and $\Qat{\alpha}{V} = \Patp{j}$ for all $\alpha \in (\alphap{j+1}, \alphap{j}] $ and $j \in \Set{0,\dotsc, p-1}$ \cite{MinAveCost,Narayanan1991PLP}. The partition chain in \eqref{eq:PSPPat}, together with the corresponding critical values $\alphap{j}$, is called the \emph{Principal Sequence of Partitions (PSP)} of the ground set $V$.

The first critical point of the PSP provides the solution to the minimum sum-rate problem \cite[Corollary A.3]{Ding2018IT}: $\RACO(V) = \alphap{1}$ for the asymptotic model and $\RNCO(V) = \lceil \alphap{1} \rceil$ for the non-asymptotic model. The corresponding partition $\Patp{1}$, called the \emph{fundamental partition}, equals to the finest maximizer of \eqref{eq:MinSumRateACOPat}.

\subsubsection{$\CoordSatCapFus$ Algorithm}

All of the existing algorithms in \cite{Ding2018IT,CourtIT2014,MiloIT2016} for solving the minimum sum-rate problem in \eqref{eq:MinSumRatePat} run a subroutine that determines the minimum and/or the finest minimizer of the Dilworth truncation~\eqref{eq:Dilworth} for a given value of $\alpha$. This subroutine is outlined by the $\CoordSatCapFus$ algorithm in Algorithm~\ref{algo:CoordSatCapFus}.
The idea is to keep increasing each dimension of a rate vector $\rv_{\alpha,V}$ in the \emph{submodular polyhedron} of $\Fu{\alpha}$
    $$ P(\Fu{\alpha}) = \Set{\rv_{\alpha,V} \in \Real^{|V|} \colon r_{\alpha}(X) \leq \Fu{\alpha}(X), X \subseteq V} $$
until it reaches the \emph{base polyhedron} of the Dilworth truncation\footnote{The original purpose of the $\CoordSatCapFus$ algorithm is to determine the value of $\FuHat{\alpha}(V)$ by tightening the upper bound $\Fu{\alpha}(X)$ in $P(\Fu{\alpha})$. See~\cite[Appendix~B]{Ding2018IT}, Also note that, since $\FuHat{\alpha}(X) \leq \Fu{\alpha}(X),\forall X \subseteq V$, $B(\FuHat{\alpha})$ and $B(\Fu{\alpha})$ are not equivalent in general.} $\FuHat{\alpha}$
    $$ B(\FuHat{\alpha}) = \Set{\rv_{\alpha,V} \in P(\Fu{\alpha}) \colon r_{\alpha}(V) = \FuHat{\alpha}(V)}. $$
Here, $\rv_{\alpha,V} = (r_{\alpha,i} \colon i \in V)$ is a $|V|$-dimension rate vector that is parameterized by the input minimum sum-rate estimate $\alpha$ and $r_\alpha (X) = \sum_{i \in X} r_{\alpha,i}, \forall X \subseteq V$ is the sum-rate function of this rate vector.
The amount of the rate increment is determined by the minimization of the set function
\begin{equation} \label{eq:FusFunc}
    \FuU{\alpha}(\TX) = \Fu{\alpha}(\TX) - r_{\alpha}(\TX), \quad \forall \X \subseteq \Qat{\alpha}{V_i}.
\end{equation}
where $\Qat{\alpha}{V_i} \in \Pi(V_i)$ is a partition of $V_i$ that is iteratively updated in Algorithm~\ref{algo:CoordSatCapFus}.
Here, we use the notation $\Qat{\alpha}{V_i}$ because we will show in Section ~\ref{subsec:PrePar} that $\Qat{\alpha}{V_i} = \bigwedge \argmin_{\Pat \in \Pi(V_i)} \Fu{\alpha}[\Pat]$ after step~\ref{step:Updates} for all $i$.
The reason for considering the function $\FuU{\alpha}(\TX)$ is the min-max relationship~\cite[Section~2.3]{Fujishige2005}~\cite[Lemmas~22 and 23]{Ding2018IT}\footnote{ Equation~\eqref{eq:MinMax} is the max-min theorem in~\cite[Section~2.3]{Fujishige2005} that holds for all $i' \in V$ and $\rv_{\alpha,V} \in P(\Fu{\alpha})$; Equation~\eqref{eq:Fusion} is proved by \cite[Lemmas~22 and 23]{Ding2018IT} for Algorithm~\ref{algo:CoordSatCapFus}, which states that the minimum of~\eqref{eq:MinMax} at the $i$th iteration can be searched over the subset $V_i$, or more specifically $\Qat{\alpha}{V_i}$, a partition of $V_i$.}: for each $\alpha$,
    \begin{subequations}
        \begin{align}
            & \max \Set{\xi \colon \rv_{\alpha,V} + \xi \chi_{i'} \in P(\Fu{\alpha})}   \label{eq:SatCap}\\
            & \qquad  = \min \Set{ \Fu{\alpha}(X) \colon i' \in X \subseteq V } \label{eq:MinMax} \\
            & \qquad  = \min\Set{ \FuU{\alpha}(\TX) \colon \Set{i'} \in \X \subseteq \Qat{\alpha}{V_i}}, \quad  \forall i' \in V_i, \label{eq:Fusion}
        \end{align}
    \end{subequations}
where \eqref{eq:Fusion} is the minimization problem in step~\ref{step:MinFus} of Algorithm~\ref{algo:CoordSatCapFus} and is a SFM problem \cite[Section~V-B]{Ding2018IT}.
The maximum of \eqref{eq:SatCap} is called the \emph{saturation capacity}.
At the end of Algorithm~\ref{algo:CoordSatCapFus}, the partition $\Qat{\alpha}{V}$ is updated to the finest minimizer of $\min_{\Pat\in \Pi(V)} \Fu{\alpha}[\Pat]$ \cite[Section~V-B]{Ding2018IT} so that $r_{\alpha}(V) = \FuHat{\alpha}(V)$.
For the CO problem, the input function $f$ in Algorithm~\ref{algo:CoordSatCapFus} refers to the entropy function $H$. But, the $\CoordSatCapFus$ algorithm generally applies to any submodular function $f$.\footnote{In this cases, $\Fu{\alpha}$ is defined as $\Fu{\alpha}(X) = \alpha - f(V) +f(X), \forall X \subseteq V$. }
In Section~\ref{sec:Relation}, we show another example of $f$, the cut function of a graph.

For solving the minimum sum-rate problem, the MDA algorithm proposed in \cite[Algorithm ~1]{Ding2018IT} utilizes the outputs of the $\CoordSatCapFus$ algorithm and the properties of the PSP in Lemma~\ref{lemma:AlphaAdapt} in Appendix~\ref{app:AlphaAdapt} to update $\alpha$ towards $\RACO(V)$. Due to the equivalence $B(\FuHat{\alpha}) = \Set{\rv_V \in \RRCO(V) \colon r(V) = \alpha}$ for all $\alpha \geq \RACO(V)$ \cite[Section~III-B and Theorem~4]{Ding2018IT}, in the final call of the $\CoordSatCapFus$ with the input $\alpha = \RACO(V)$, an optimal rate vector $\rv_{\RACO(V),V} \in B(\FuHat{\RACO(V)}) = \RRACO(V)$ is also returned.
For the non-asymptotic model, an optimal rate vector $\rv_{\RNCO(V),V} \in B(\FuHat{\RNCO(V)}) \cap \Z^{|V|} = \RRNCO(V)$ can be determined by running the $\CoordSatCapFus$ algorithm with the input $\alpha = \RNCO(V) = \lceil \RACO(V) \rceil$.

For the input $\alpha = \RACO(V)$, the $\CoordSatCapFus$ algorithm also outputs the fundamental partition $ \Qat{\RACO(V)}{V} = \Patp{1}$. This is an important parameter in CCDE in that it is the least common multiple (LCM) of $\rv_{\RACO(V),V}$ \cite[Corollary ~28]{Ding2018IT}, i.e., by letting each packet be broken into $|\Patp{1}| - 1$ chunks, the optimal rate vector $\rv_{\RACO(V),V}$ is implementable based on linear codes, which saves the overall transmission rates by no more than $1$ from the optimal rate vector $\rv_{\RNCO(V),V} \in \RRNCO(V)$.

        \begin{algorithm} [t]
	       \label{algo:CoordSatCapFus}
	       \small
	       \SetAlgoLined
	       \SetKwInOut{Input}{input}\SetKwInOut{Output}{output}
	       \SetKwFor{For}{for}{do}{endfor}
            \SetKwRepeat{Repeat}{repeat}{until}
            \SetKwIF{If}{ElseIf}{Else}{if}{then}{else if}{else}{endif}
	       \BlankLine
           \Input{$\alpha$, $f$, $V$ and $\Phi$}
	       \Output{$\rv_{\alpha,V} \in B(\FuHat{\alpha})$ and $\Qat{\alpha}{V} = \bigwedge \argmin_{\Pat\in \Pi(V)} \Fu{\alpha}[\Pat]$ }
	       \BlankLine
            Let $\rv_{\alpha,V} \coloneqq (\alpha - H(V)) \chi_V$ so that $\rv_{\alpha,V} \in P(\Fu{\alpha})$\;
            Initiate $ r_{\alpha,\phi_1} \coloneqq \Fu{\alpha}(\Set{\phi_1})$ and $\Qat{\alpha}{V_1} \coloneqq \Set{\Set{\phi_1}}$\;
            \For{$i=2$ \emph{\KwTo} $|V|$}{
                $\Qat{\alpha}{V_i} \coloneqq \Qat{\alpha}{V_{i-1}} \sqcup \Set{\Set{\phi_i}}$ \label{step:PatIni} \;
                $\U{\alpha}{V_i} \coloneqq \bigcap \argmin\Set{ \FuU{\alpha}(\TX) \colon \Set{\phi_i} \in \X \subseteq \Qat{\alpha}{V_i}}$\; \label{step:MinFus}
                Update $\rv_{\alpha,V}$ and $\Qat{\alpha}{V_i}$: \label{step:Updates}
                \begin{equation}
                    \begin{aligned}
                        \rv_{\alpha,V} &\coloneqq \rv_{\alpha,V} + \FuU{\alpha}(\TU{\alpha}{V_i}) \chi_{\phi_i}; \\
                        \Qat{\alpha}{V_i} &\coloneqq (\Qat{\alpha}{V_i} \setminus \U{\alpha}{V_i}) \sqcup \Set{ \TU{\alpha}{V_i} };
                    \end{aligned} \nonumber
                \end{equation}
            }
            \Return $\rv_{\alpha,V}$ and $\Qat{\alpha}{V}$\;
	   \caption{$\CoordSatCapFus$ Algorithm \cite[Algorithm~3]{Ding2018IT}}
	   \end{algorithm}

\section{Parametric Approach}
\label{sec:ParAlgo}

While the $\CoordSatCapFus$ algorithm determines the Dilworth truncation $\FuHat{\alpha}(V)$ for only one value of $\alpha$, we reveal the structural properties of the partition $\Qat{\alpha}{V_i}$ and the rate vector $\rv_{\alpha,V}$ in $\alpha$ and show that the objective function $\FuU{\alpha}(\TX)$ in the SFM problem \eqref{eq:Fusion} exhibits the strict strong map property in $\alpha$.
A parametric ($\PAR$) algorithm is then proposed, which utilizes this strong map property to obtain the minimizer of \eqref{eq:Fusion} for all $\alpha$ so that each iteration $i$ determines $\Qat{\alpha}{V_i}$ and $\rv_{\alpha,V}$, in particular its reduction $\rv_{\alpha,V_i}$ on $V_i$, for all values of the minimum sum-rate estimate $\alpha$.
Also, by choosing a proper linear ordering $\Phi$, the optimal rate vector $\rv_{\alpha,V}$ returned by the $\CoordSatCapFus$ algorithm for both asymptotic and non-asymptotic models minimizes a weighted minimum sum-rate objective function.
We show in Section~\ref{sec:Complexity} that this PAR algorithm reduces the computational complexity for solving the minimum sum-rate problem in both asymptotic and non-asymptotic models and allows distributed computation.
Note that, in this paper, when we say for all $\alpha$, we mean for all $\alpha \in [0, H(V)]$ since the minimum sum-rates, $ \RACO(V)$ and $\RNCO(V)$, must take values in $[0, H(V)]$.

\subsection{Observations}
\label{subsec:PrePar}

Observing the values of $\Qat{\alpha}{V_i}$ and $\rv_{\alpha,V_i}$ in $\alpha$ in the $\CoordSatCapFus$ algorithm as the iteration index $i$ grows, we have the following result.
\begin{proposition} \label{prop:preamble}
    After step~\ref{step:Updates} in each iteration $i$ of Algorithm~\ref{algo:CoordSatCapFus}, $\Qat{\alpha}{V_i} = \bigwedge \argmin_{\Pat \in \Pi(V_i)} \Fu{\alpha}[\Pat]$ and $\rv_{\alpha,V_i} \in B(\FuHat{\alpha}^{V_i})$ for all $\alpha$. \hfill \IEEEQED
\end{proposition}
The proof is omitted since it is a direct result that $\Qat{\alpha}{V_i} = \bigwedge \argmin_{\Pat \in \Pi(V_i)} \Fu{\alpha}[\Pat]$ and $\rv_{\alpha,V_i} \in B(\FuHat{\alpha}^{V_i})$ are returned by the call $\CoordSatCapFus(\alpha,H,V_i,\Phi)$.
Note that, due to the equivalence $B(\FuHat{\alpha}^{V_i}) = \Set{\rv_{V_i} \in \RRCO(V_i) \colon r(V_i) = \alpha}$, we must have the reduction of the rate vector $\rv_{\alpha,V}$ on $V_i$ being $\rv_{\alpha,V_i} \in B(\FuHat{\alpha}^{V_i})$ for all $\alpha$ after step~\ref{step:Updates}.
According to Proposition~\ref{prop:preamble}, $\Qat{\alpha}{V_i}$ for all $\alpha$ is again characterized by the PSP of $V_i$ with the number of critical points bounded by $|V_i|$.
That is, the partition $\Qat{\alpha}{V_i}$ and $\rv_{\alpha,V_i}$ are segmented in $\alpha$ and determine the solution to the minimum sum-rate problem in a subsystem $V_i$.
This fact will be utilized in Section~\ref{sec:Distr} to propose a distributed algorithm for solving the CO problem and in Part II of this paper~\cite{DingITSO2019} to solve the successive omniscience problem.

\begin{example} \label{ex:main}
    Consider a $5$-user system with
    \begin{equation}
        \begin{aligned}
            \RZ{1} & = (\RW{b},\RW{c},\RW{d},\RW{h},\RW{i}),   \\
            \RZ{2} & = (\RW{e},\RW{f},\RW{h},\RW{i}),   \\
            \RZ{3} & = (\RW{b},\RW{c},\RW{e},\RW{j}), \\
            \RZ{4} & = (\RW{a},\RW{b},\RW{c},\RW{d},\RW{f},\RW{g},\RW{i},\RW{j}),  \\
            \RZ{5} & = (\RW{a},\RW{b},\RW{c},\RW{f},\RW{i},\RW{j}),
        \end{aligned}  \nonumber
    \end{equation}
    where each $\RW{m}$ is an independent uniformly distributed random bit.

    Choose the linear ordering $\Phi = (4,5,2,3,1)$.
    By setting $\alpha = 3$, we call $\CoordSatCapFus(\alpha,H,V,\Phi)$ . We initiate $\rv_{3,V} = (\alpha - H(V)) \chi_V = (-7,\dotsc,-7)$, update $r_{3,4} = \Fu{3}(\Set{4}) = 1$ and assign $\Qat{3}{V_1} = \Set{\Set{4}}$. For $i  = 2$ and $\phi_2=5$, consider the minimization problem $\min\Set{ \FuU{3}(\TX) \colon \Set{5} \in \X \subseteq \Qat{3}{V_2}}$ where $\Qat{3}{V_2} = \Set{\Set{4},\Set{5}}$.
    We get the minimal minimizer $\U{3}{V_2} = \Set{\Set{5}}$ and do the updates $r_{3,2} = -7 + \FuU{3}(\TU{3}{V_2}) = -1$ and $\Qat{3}{V_2} = \big( \Set{\Set{4},\Set{5}} \setminus \U{3}{V_2} \big) \sqcup \Set{\TU{3}{V_2}} = \Set{\Set{4},\Set{5}}$. In the same way, one can continue the rest of iterations. However, to show an example of Proposition~\ref{prop:preamble}, we consider another value of $\alpha$ as follows.

    By setting $\alpha = 6$, we call $\CoordSatCapFus(\alpha,H,V,\Phi)$. We initiate $\rv_{6,V} = (\alpha - H(V)) \chi_V = (-4,\dotsc,-4)$ and set $r_{6,4} = \Fu{6}(\Set{4}) = 4$ and $\Qat{6}{V_1} = \Set{\Set{4}}$. We have $\U{6}{V_2} = \bigcap \argmin\Set{ \FuU{6}(\TX) \colon \Set{5} \in \X \subseteq \Qat{6}{V_2}} = \Set{\Set{4},\Set{5}} $ and the updates $r_{6,5} = -4 + \FuU{6}(\TU{6}{V_2}) = 0$ and $\Qat{6}{V_2} = \big( \Set{\Set{4},\Set{5}} \setminus \U{6}{V_2} \big) \sqcup \Set{\TU{6}{V_2}} = \Set{\Set{4,5}}$.

    One can verify that $\Set{\Set{4},\Set{5}} = \bigwedge \argmin_{\Pat \in \Pi(V_2)} \Fu{3}[\Pat] $ and $\Set{\Set{4,5}} = \bigwedge \argmin_{\Pat \in \Pi(V_2)} \Fu{6}[\Pat] $. In fact, repeating the above procedure for all $\alpha$, we have the piecewise linear $\rv_{\alpha,\Set{4,5}}$ and segmented $\Qat{\alpha}{V_2}$ as
    \begin{equation} \label{eq:ExQatR}
        \begin{aligned}
            & r_{\alpha,4} = \alpha-2, \quad \forall \alpha \in [0,10], \\
            & r_{\alpha,5} =  \begin{cases}
                                    \alpha-4 & \alpha \in [0,4], \\
                                    0 & \alpha \in (4,10],
                                \end{cases} \\
            & \Qat{\alpha}{V_2} = \begin{cases}
                                    \Set{\Set{4},\Set{5}} & \alpha \in [0,4], \\
                                    \Set{\Set{4,5}} & \alpha \in (4,10],
                                \end{cases}
        \end{aligned}
    \end{equation}
    because of the segmented
    \begin{equation} \label{eq:ExU}
        \TU{\alpha}{V_2} = \begin{cases}
                             \Set{5} & \alpha \in [0,4],\\
                             \Set{4,5} & \alpha \in (4,10].
                         \end{cases}
    \end{equation}
    Note that the function $\FuU{\alpha}$ is segmented. For example, for $i  = 3$ and $\phi_3=2$, the function $\FuU{\alpha}$ defined on $\Qat{\alpha}{V_{2}} \sqcup \Set{\Set{2}}$ differs in two segments: for $\alpha \in [0,4]$, $\FuU{\alpha}$ takes values on $\emptyset$, $\Set{2}$, $\Set{4,5}$ and $\Set{2,4,5}$ only; for $\alpha \in (4,10]$, $\FuU{\alpha}$ takes values on all subsets in the power set $2^{\Set{2,4,5}}$.
\end{example}

Proposition~\ref{prop:preamble} suggests that we could obtain $\rv_{\alpha,V_i}$ and $\Qat{\alpha}{V_i}$ for all values of $\alpha$ in each iteration of the $\CoordSatCapFus$ algorithm. To do so, it is essential to discuss how to efficiently determine $\TU{\alpha}{i}$ for all $\alpha$.
It should be noted that we automatically know $\U{\alpha}{V_i}$ if $\TU{\alpha}{V_i}$ is obtained in that $\U{\alpha}{V_i} = \Set{ C \in \Qat{\alpha}{V_{i}} \colon C \subseteq \TU{\alpha}{V_i}} = \ASet{\TU{\alpha}{V_i}}{\Qat{\alpha}{V_i}}$.\footnote{This means that $\U{\alpha}{V_i}$ is the decomposition of $\TU{\alpha}{V_i}$ by $\Qat{\alpha}{V_{i}}$. Here, we should use the value of $\Qat{\alpha}{V_i}$ in the minimization problem $\min\Set{ \FuU{\alpha}(\TX) \colon \Set{i} \in \X \subseteq \Qat{\alpha}{V_i}}$ before the updates in step~\ref{step:Updates}. }
For example, for $\TU{\alpha}{V_2}$ in \eqref{eq:ExU},
    \begin{equation} \label{eq:ExU2}
        \U{\alpha}{V_2} = \begin{cases}
                             \Set{\Set{5}} & \alpha \in [0,4],\\
                             \Set{\Set{4},\Set{5}} & \alpha \in (4,10].
                         \end{cases}
    \end{equation}
We derive the following structural results on $\Qat{\alpha}{V_i}$ and $\rv_{\alpha,V}$.

\begin{lemma}[essential properties] \label{lemma:EssProp}
    At the end of each iteration $i$ of Algorithm~\ref{algo:CoordSatCapFus}, the rate vector $\rv_{\alpha,V} \in P(\Fu{\alpha})$, where $ P(\Fu{\alpha}) = P(\FuHat{\alpha})$, and followings hold for all $\alpha$:
    \begin{enumerate}[(a)]
        \item $r_\alpha(V_i) = r_\alpha [\Qat{\alpha}{V_i}] = \Fu{\alpha}[\Qat{\alpha}{V_i}]= \FuHat{\alpha} (V_i)$;
        \item $r_\alpha(\TX) = r_\alpha[\X] = \Fu{\alpha}[\X] = \FuHat{\alpha}(\TX)$ for all $ \X \subseteq \Qat{\alpha}{V_i}$,
    \end{enumerate}
    where $r_{\alpha}[\X] = \sum_{C \in \X} r_{\alpha}(C)$.
    \begin{enumerate}[(c)]
        \item For all $\alpha < \alpha'$, $\Qat{\alpha}{V_i} \preceq \Qat{\alpha'}{V_i}$ and, for all $\X \subseteq \Qat{\alpha}{V_i}$ and $\X' \subseteq \Qat{\alpha'}{V_i}$ such that $\TX = \TX'$,
        $$ r_{\alpha}[\X] = \FuHat{\alpha}(\TX) < \FuHat{\alpha'}(\TX') = r_{\alpha'}[\X'] . \IEEEQEDhereeqn$$
    \end{enumerate}
\end{lemma}
The proof of Lemma~\ref{lemma:EssProp} is in Appendix~\ref{app:lemma:EssProp}. We call Lemma~\ref{lemma:EssProp}(a) to (c) the essential properties since they hold the strict strong map property in the Theorem~\ref{theo:StrongMap} in Section~\ref{subsec:StrMap}, the main theorem that ensures the validity of the $\PAR$ algorithm. In addition, in Part II \cite{DingITSO2019}, we show that the monotonicity of the sum-rate in Lemma~\ref{lemma:EssProp}(c) also guarantees the feasibility of a multi-stage SO. 

\subsection{Strong Map Property}
\label{subsec:StrMap}

Since $\Qat{\alpha}{V_i} = \bigwedge \argmin_{\Pat \in \Pi(V_i)} \Fu{\alpha}[\Pat]$ after
step~\ref{step:Updates} of Algorithm~\ref{algo:CoordSatCapFus} according to Proposition~\ref{prop:preamble}, $\Qat{\alpha}{V_i}$ satisfies the properties of the PSP in Section~\ref{subsec:PSP}, i.e., the partition $\Qat{\alpha}{V_i}$ gets monotonically coarser as $\alpha$ increases and is segmented by a finite number of critical points.
Recall that $\Qat{\alpha}{V_i}$ is updated by $\U{\alpha}{V_i}$ in step~\ref{step:Updates}. Then, we must have $\TU{\alpha}{V_i}$ segmented in $\alpha$ and the size of $\TU{\alpha}{V_i}$ increase in $\alpha$.
This can be justified by the strong map property of the function $\FuU{\alpha}$, which also states that all critical points that characterize the segmented $\TU{\alpha}{V_i}$ can be determined by the parametric submodular function minimization (PSFM) algorithm.

\begin{definition}[strong map {\cite[Section~4.1]{Fujishige2009PP}}] \label{def:StrMap}
    For two distributive lattices $\D_1,\D_2 \subseteq 2^V$,\footnote{A group of sets $\D$ form a distributive lattice if, for all $X,Y \in \D$, $X \cap Y \in \D$ and $X \cup Y \in \D$ \cite[Section~3.2]{Fujishige2005}.}
    and submodular functions $h_1 \colon \D_1 \mapsto \Real$ and $h_2 \colon \D_2 \mapsto \Real$, $h_1$ and $h_2$ form a strong map, denoted by $h_1 \rightarrow h_2$, if
    \begin{equation}\label{eq:StrongMap}
        h_1(Y) - h_1(X) \geq h_2(Y) - h_2(X)
    \end{equation}
    for all $X,Y \in \D_1 \cap \D_2$ such that $X \subseteq Y$. The strong map is strict, denoted by $h_1 \twoheadrightarrow h_2$, if $h_1(Y) - h_1(X) > h_2(Y) - h_2(X)$ for all $X \subsetneq Y$.
\end{definition}

\begin{theorem} \label{theo:StrongMap}
    In each iteration $i$ of Algorithm~\ref{algo:CoordSatCapFus}, $\FuU{\alpha}$ forms a \textbf{strict strong map} in $\alpha$:
        $$ \FuU{\alpha} \twoheadrightarrow \FuU{\alpha'}, \quad \forall \alpha, \alpha' \colon \alpha < \alpha'. $$
\end{theorem}
\begin{IEEEproof}
    For any $\X \subseteq \Qat{\alpha}{V_i}$ and $\Y \subseteq \Qat{\alpha'}{V_i}$ such that $i \in \TX \subseteq \TY$ and $\FuU{\alpha}$ and $\FuU{\alpha'}$ are both defined on $\TX$ and $\TY$, we have $i \notin \TY \setminus \TX$. Also, there exist $\M \subseteq \Qat{\alpha}{V_i}$ and $\N \subseteq \Qat{\alpha'}{V_i}$ (with $\M \preceq \N$) such that $\TM = \TN = \TY \setminus \TX$. According to Lemma~\ref{lemma:EssProp}(b) and (c), $r_{\alpha}(\TY \setminus \TX) = \Fu{\alpha}[\M] = \FuHat{\alpha}(\TY \setminus \TX)$ and $r_{\alpha'}(\TY \setminus \TX) = \Fu{\alpha'}[\N] = \FuHat{\alpha'}(\TY \setminus \TX)$.
    Then,
    \begin{equation}
        \begin{aligned}
            & \FuU{\alpha}(\TY) - \FuU{\alpha}(\TX) - \FuU{\alpha'}(\TY) + \FuU{\alpha'}(\TX) \\
            &\qquad\qquad\qquad = r_{\alpha'}(\TY \setminus \TX) - r_{\alpha}(\TY \setminus \TX)\\
            &\qquad\qquad\qquad = \begin{cases}
                    0 & \TX = \TY, \\
                    \FuHat{\alpha'} (\TY \setminus \TX)  - \FuHat{\alpha} (\TY \setminus \TX) & \TX \subsetneq \TY,
                \end{cases}
        \end{aligned} \nonumber
    \end{equation}
    where $\FuHat{\alpha'} (\TY \setminus \TX)  - \FuHat{\alpha} (\TY \setminus \TX) > 0$ for all $\alpha$ and $\alpha'$ such that $\alpha < \alpha'$ based on Lemma~\ref{lemma:EssProp}(c). This proves the theorem according to Definition~\ref{def:StrMap}.
\end{IEEEproof}

The strict strong map property directly leads to the structural property of $\TU{\alpha}{V_i}$ in $\alpha$.

\begin{lemma}{\cite[Theorems~26 to 28]{Fujishige2009PP}} \label{lemma:PP}
    In each iteration $i$ of Algorithm~\ref{algo:CoordSatCapFus}, the minimal minimizer $\U{\alpha}{V_i}$ of $\min\Set{ \FuU{\alpha}(\TX) \colon \Set{\phi_i} \in \X \subseteq \Qat{\alpha}{V_i}}$ satisfies $\TU{\alpha}{V_i} \subseteq \TU{\alpha'}{V_i}$ for all $\alpha < \alpha'$. In addition, $\TU{\alpha}{V_i} $
    for all $\alpha$ is fully characterized by $q < |V_i| - 1$ critical points
    $$ 0 \leq \alpha_q < \dotsc < \alpha_1 < \alpha_0 = H(V) $$
    and the corresponding minimal minimizer $\TUp{j} = \TU{\alpha_j}{V_i}$ for all $j \in \Set{0,\dotsc,q}$ forms a set chain
    $$ \Set{\phi_i} = \TUp{q} \subsetneq  \dotsc \subsetneq \TUp{1} \subsetneq \TUp{0} = V_i $$
    and $\TU{\alpha}{V_i} = \TUp{q} = \Set{\phi_i}$ for all $\alpha \in [0,\alpha_q]$ and $\TU{\alpha}{V_i} = \TUp{j}$ for all $\alpha \in (\alpha_{j+1}, \alpha_j]$ such that $j \in \Set{0,\dotsc,q-1}$.\footnote{It should be noted that the value of $\alphap{j}$'s in the PSP and $\alpha_{j}$'s in Lemma~\ref{lemma:PP} do not necessarily coincide. The critical points $\alpha_j$'s for $\min\Set{ \FuU{\alpha}(\TX) \colon \Set{\phi_i} \in \X \subseteq \Qat{\alpha}{V_i}}$ for each iteration $i$ also vary with the linear ordering $\Phi$. } \hfill\IEEEQED
\end{lemma}

\begin{example} \label{ex:PP}
    In Example~\ref{ex:main}, we have $\TU{\alpha}{V_2}$ in \eqref{eq:ExU} characterized by the critical points $\alpha_1 = 4$ and $\alpha_0 = H(V) = 10$ with $\TUp{1} = \Set{5}$ and $\TUp{0} = \Set{4,5}$ such that
    $ \Set{5} = \TUp{1} \subsetneq  \TUp{0} = V_2 $. So, $\TU{\alpha}{V_2} = \TUp{1}$ for $\alpha \in [0,\alpha_1]$ and $\TU{\alpha}{V_2} = \TUp{0}$ for $\alpha \in (\alpha_1,\alpha_0]$.

    We continue the procedure in Example~\ref{ex:main} for $i = 3$ and $\phi_3 = 2$ by considering the problem $\min\Set{ \FuU{\alpha}(\TX) \colon \Set{2} \in \X \subseteq \Qat{\alpha}{V_2} \sqcup \Set{\Set{2}}}$ where $\Qat{\alpha}{V_2}$ and $\rv_{\alpha,V_2}$ are in \eqref{eq:ExQatR}. We have
    \begin{equation}  \label{eq:UpV3}
        \TU{\alpha}{V_3} = \begin{cases}
                             \Set{2} & \alpha \in [0,8],\\
                             \Set{2,4,5} & \alpha \in (8,10]
                         \end{cases}
    \end{equation}
    that is determined by the critical points $\alpha_1 = 8$ and $\alpha_0 = H(V) = 10$ with $\TUp{1} = \Set{2}$ and $\TUp{0} = \Set{2,4,5}$ such that
    $ \Set{2} = \TUp{1} \subsetneq  \TUp{0} = V_3 = \Set{2,4,5}$.
    After the updates in step~\ref{step:Updates}, we have
        \begin{equation}  \label{eq:UpdateV3}
        \begin{aligned}
            & \rv_{\alpha,V} = \begin{cases}
                                    (\alpha-10,\alpha-6, \alpha-10, \alpha-2, \alpha-4) & \alpha \in [0,4], \\
                                    (\alpha-10, \alpha-6, \alpha-10, \alpha-2, 0) & \alpha \in (4,8], \\
                                    (\alpha-10, 2, \alpha-10, \alpha-2,0) & \alpha \in (8,10],
                                \end{cases} \\
            & \Qat{\alpha}{V_3} = \begin{cases}
                                    \Set{\Set{2},\Set{4},\Set{5}} & \alpha \in [0,4], \\
                                    \Set{\Set{4,5},\Set{2}} & \alpha \in (4,8], \\
                                    \Set{\Set{2,4,5}} & \alpha \in (8,10].
                                \end{cases}
        \end{aligned}
        \end{equation}
\end{example}

\subsection{Parametric Algorithm}

Lemma~\ref{lemma:PP} directly leads to the $\PAR$ algorithm in Algorithm~\ref{algo:ParAlgo}, where the values of $\Qat{\alpha}{V_i}$ and $\rv_{\alpha,V_i}$ are determined for all $\alpha$ in each iteration $i$. We call Algorithm~\ref{algo:ParAlgo} a parametric algorithm since the variables $\TU{\alpha}{V_i}$, $\Qat{\alpha}{V_i}$ and $\rv_{\alpha,V_i}$ are parameterized by the minimum sum-rate estimate $\alpha$.
For the minimum sum-rate problems in~\eqref{eq:MinSumRatePat}, the input linear ordering $\Phi$ can be arbitrarily chosen. In Section~\ref{sec:WeightedSum}, we show how to choose $\Phi$ to minimize a weighted sum-rate problem.

      \begin{algorithm} [t]
	       \label{algo:ParAlgo}
	       \small
	       \SetAlgoLined
	       \SetKwInOut{Input}{input}\SetKwInOut{Output}{output}
	       \SetKwFor{For}{for}{do}{endfor}
            \SetKwRepeat{Repeat}{repeat}{until}
            \SetKwIF{If}{ElseIf}{Else}{if}{then}{else if}{else}{endif}
	       \BlankLine
           \Input{$f$, $V$ and $\Phi$}
	       \Output{segmented variables $\rv_{\alpha,V} \in B(\FuHat{\alpha})$ and $\Qat{\alpha}{V} = \bigwedge \argmin_{\Pat\in \Pi(V)} \Fu{\alpha}[\Pat]$ for all $\alpha$}
	       \BlankLine
            $\rv_{\alpha,V} \coloneqq (\alpha - H(V)) \chi_V$ for all $\alpha$\;
            $ r_{\alpha,\phi_1} \coloneqq \Fu{\alpha}(\Set{\phi_1})$ and $\Qat{\alpha}{V_1} \coloneqq \Set{\Set{\phi_1}}$ for all $\alpha$\;
            \For{$i=2$ \emph{\KwTo} $|V|$}{
                $\Qat{\alpha}{V_i} \coloneqq \Qat{\alpha}{V_{i-1}} \sqcup \Set{\Set{\phi_i}}$ for all $\alpha$\;
                Obtain the critical points $\Set{\alpha_j \colon j \in \Set{0,\dotsc,q}}$ and $\Set{\TUp{j} \colon j \in \Set{0,\dotsc,q}}$ that determine the minimal minimizer $\U{\alpha}{V_i}$ of $\min\Set{ \FuU{\alpha}(\TX) \colon \Set{\phi_i} \in \X \subseteq \Qat{\alpha}{V_i}}$ for all $\alpha$ by the StrMap algorithm in Algorithm~\ref{algo:StrMap}\label{step:PP}\;
                Let $\Gamma_j \coloneqq (\alpha_{j+1},\alpha_{j}]$ for all $j \in \Set{0,\dotsc,q-1}$ and $\Gamma_q \coloneqq [0,\alpha_q]$. For each $j \in \Set{0,\dotsc,q}$, update $\rv_V$ and $\Qat{\alpha}{V_i}$ by \label{step:UpdatesPar}
                \begin{equation}
                    \begin{aligned}
                        \rv_{\alpha,V} &\coloneqq \rv_{\alpha,V} + \FuU{\alpha}(\TUp{j}) \chi_{\phi_i}; \\
                        \Qat{\alpha}{V_i} &\coloneqq (\Qat{\alpha}{V_i} \setminus \Up{j}) \sqcup \Set{ \TUp{j} };
                    \end{aligned} \nonumber
                \end{equation}
                for all $\alpha \in \Gamma_j$;
            }
            \Return $\rv_V$ and $\Qat{\alpha}{V}$ for all $\alpha$\;
	   \caption{Parametric (PAR) Algorithm}
	   \end{algorithm}

\begin{figure*}[t]
	\centering
    \subfigure[$\FuHat{\alpha}(V_1)$ and $\Qat{\alpha}{V_1}$]{\scalebox{0.6}{
%
%
\definecolor{mycolor1}{rgb}{1,0,1}%

\begin{tikzpicture}[
every pin/.style={fill=yellow!50!white,rectangle,rounded corners=3pt,font=\tiny},
every pin edge/.style={<-}]

\begin{axis}[%
width=2.8in,
height=0.8in,
scale only axis,
xmin=0,
xmax=10,
xlabel={\Large $\alpha$},
xmajorgrids,
ymin=-5,
ymax=10,
ylabel={\large $\FuHat{\alpha}(V_1)$},
ymajorgrids,
legend style={at={(0.65,0.05)},anchor=south west,draw=black,fill=white,legend cell align=left}
]

\addplot [
color=blue,
solid,
line width=1.5pt,
mark=asterisk,
mark options={solid}
]
table[row sep=crcr]{
10 8\\
0 -2\\
};

\addplot[area legend,solid,fill=blue!20,opacity=4.000000e-01]coordinates {
(10,10)
(10,-8)
(0,-8)
(0,10)
};
\node at (axis cs:5,0) {\textcolor{blue}{\large $\Set{\Set{4}}$}};


\end{axis}
\end{tikzpicture}%
}}
    \subfigure[$\FuHat{\alpha}(V_2)$ and $\Qat{\alpha}{V_2}$]{\scalebox{0.6}{
%
%
\definecolor{mycolor1}{rgb}{1,0,1}%

\begin{tikzpicture}[
every pin/.style={fill=yellow!50!white,rectangle,rounded corners=3pt,font=\tiny},
every pin edge/.style={<-}]

\begin{axis}[%
width=2.8in,
height=0.8in,
scale only axis,
xmin=0,
xmax=10,
xlabel={\Large $\alpha$},
xmajorgrids,
ymin=-8,
ymax=10,
ylabel={\large $\FuHat{\alpha}(V_2)$},
ymajorgrids,
legend style={at={(0.65,0.05)},anchor=south west,draw=black,fill=white,legend cell align=left}
]

\addplot [
color=blue,
solid,
line width=1.5pt,
mark=asterisk,
mark options={solid}
]
table[row sep=crcr]{
10 8\\
4 2\\
};

\addplot[area legend,solid,fill=blue!20,opacity=4.000000e-01]coordinates {
(10,10)
(10,-8)
(4,-8)
(4,10)
};
\node at (axis cs:7,0) {\textcolor{blue}{\large $\Set{\Set{4,5}}$}};

\addplot [
color=orange,
solid,
line width=1.5pt,
mark=asterisk,
mark options={solid}
]
table[row sep=crcr]{
4 2\\
0 -6 \\
};

\addplot[area legend,solid,fill=orange!20,opacity=4.000000e-01]coordinates {
(4,10)
(4,-8)
(0,-8)
(0,10)
};
\node at (axis cs:2,5) {\textcolor{orange}{\large $\Set{\Set{4},\Set{5}}$}};



\end{axis}
\end{tikzpicture}%
}}
    \subfigure[$\FuHat{\alpha}(V_3)$ and $\Qat{\alpha}{V_3}$]{\scalebox{0.6}{
%
%
\definecolor{mycolor1}{rgb}{1,0,1}%

\begin{tikzpicture}[
every pin/.style={fill=yellow!50!white,rectangle,rounded corners=3pt,font=\tiny},
every pin edge/.style={<-}]

\begin{axis}[%
width=3.5in,
height=0.8in,
scale only axis,
xmin=0,
xmax=10,
xlabel={\Large $\alpha$},
xmajorgrids,
ymin=-13,
ymax=11,
ylabel={\large $\FuHat{\alpha}(V_3)$},
ymajorgrids,
legend style={at={(0.65,0.05)},anchor=south west,draw=black,fill=white,legend cell align=left}
]

\addplot [
color=blue,
solid,
line width=1.5pt,
mark=asterisk,
mark options={solid}
]
table[row sep=crcr]{
10 10\\
8 8\\
};

\addplot[area legend,solid,fill=blue!20,opacity=4.000000e-01]coordinates {
(10,11)
(10,-13)
(8,-13)
(8,11)
};
\node at (axis cs:9,0) {\textcolor{blue}{\large $\Set{\Set{2,4,5}}$}};

\addplot [
color=purple,
solid,
line width=1.5pt,
mark=asterisk,
mark options={solid}
]
table[row sep=crcr]{
8 8\\
4 0\\
};

\addplot[area legend,solid,fill=purple!20,opacity=4.000000e-01]coordinates {
(8,11)
(8,-13)
(4,-13)
(4,11)
};
\node at (axis cs:6,-5) {\textcolor{purple}{\large $\Set{\Set{4,5},\Set{2}}$}};

\addplot [
color=orange,
solid,
line width=1.5pt,
mark=asterisk,
mark options={solid}
]
table[row sep=crcr]{
4 0\\
0 -12 \\
};

\addplot[area legend,solid,fill=orange!20,opacity=4.000000e-01]coordinates {
(4,11)
(4,-13)
(0,-13)
(0,11)
};
\node at (axis cs:2,5) {\textcolor{orange}{\large $\Set{\Set{2},\Set{4},\Set{5}}$}};



\end{axis}
\end{tikzpicture}%
}}
    \subfigure[$\FuHat{\alpha}(V_4)$ and $\Qat{\alpha}{V_4}$]{\scalebox{0.6}{
%
%
\definecolor{mycolor1}{rgb}{1,0,1}%

\begin{tikzpicture}[
every pin/.style={fill=yellow!50!white,rectangle,rounded corners=3pt,font=\tiny},
every pin edge/.style={<-}]

\begin{axis}[%
width=4in,
height=0.8in,
scale only axis,
xmin=0,
xmax=10,
xlabel={\Large $\alpha$},
xmajorgrids,
ymin=-19,
ymax=11,
ylabel={\large $\FuHat{\alpha}(V_4)$},
ymajorgrids,
legend style={at={(0.65,0.05)},anchor=south west,draw=black,fill=white,legend cell align=left}
]

\addplot [
color=blue,
solid,
line width=1.5pt,
mark=asterisk,
mark options={solid}
]
table[row sep=crcr]{
10 10\\
7 7\\
};

\addplot[area legend,solid,fill=blue!20,opacity=4.000000e-01]coordinates {
(10,11)
(10,-19)
(7,-19)
(7,11)
};
\node at (axis cs:8.5,-5) {\textcolor{blue}{\large $\Set{\Set{2,\dotsc,5}}$}};

\addplot [
color=purple,
solid,
line width=1.5pt,
mark=asterisk,
mark options={solid}
]
table[row sep=crcr]{
7 7\\
4 -2\\
};

\addplot[area legend,solid,fill=purple!20,opacity=4.000000e-01]coordinates {
(7,11)
(7,-19)
(4,-19)
(4,11)
};
\node at (axis cs:5.5,-6.5) {\textcolor{purple}{\large $\Set{\Set{4,5},\Set{2},\Set{3}}$}};

\addplot [
color=orange,
solid,
line width=1.5pt,
mark=asterisk,
mark options={solid}
]
table[row sep=crcr]{
4 -2\\
0 -18 \\
};

\addplot[area legend,solid,fill=orange!20,opacity=4.000000e-01]coordinates {
(4,11)
(4,-19)
(0,-19)
(0,11)
};
\node at (axis cs:2,2) {\textcolor{orange}{\large $\Set{\Set{2},\dotsc,\Set{5}}$}};



\end{axis}
\end{tikzpicture}%
}} \quad
    \subfigure[$\FuHat{\alpha}(V)$ and $\Qat{\alpha}{V}$, where $\Qat{\alpha}{V} = \Set{\Set{1,4,5},\Set{2},\Set{3}},\forall \alpha \in {(6,6.5]}$]{\scalebox{0.6}{
%
%
\definecolor{mycolor1}{rgb}{1,0,1}%

\begin{tikzpicture}[
every pin/.style={fill=yellow!50!white,rectangle,rounded corners=3pt,font=\tiny},
every pin edge/.style={<-}]

\begin{axis}[%
width=5.5in,
height=0.8in,
scale only axis,
xmin=0,
xmax=10,
xlabel={\Large $\alpha$},
xmajorgrids,
ymin=-23,
ymax=11,
ylabel={\large $\FuHat{\alpha}(V)$},
ymajorgrids,
legend style={at={(0.65,0.05)},anchor=south west,draw=black,fill=white,legend cell align=left}
]

\addplot [
color=blue,
solid,
line width=1.5pt,
mark=asterisk,
mark options={solid}
]
table[row sep=crcr]{
10 10\\
6.5 6.5\\
};

\addplot[area legend,solid,fill=blue!20,opacity=4.000000e-01]coordinates {
(10,11)
(10,-23)
(6.5,-23)
(6.5,11)
};
\node at (axis cs:8.25,-5) {\textcolor{blue}{\large $\Set{\Set{1,\dotsc,5}}$}};

\addplot [
color=purple,
solid,
line width=1.5pt,
mark=asterisk,
mark options={solid}
]
table[row sep=crcr]{
6.5 6.5\\
6 5\\
};

\addplot[area legend,solid,fill=purple!20,opacity=4.000000e-01]coordinates {
(6.5,11)
(6.5,-23)
(6,-23)
(6,11)
};

\addplot [
color=green,
solid,
line width=1.5pt,
mark=asterisk,
mark options={solid}
]
table[row sep=crcr]{
6 5\\
4 -3 \\
};

\addplot[area legend,solid,fill=green!20,opacity=4.000000e-01]coordinates {
(6,11)
(6,-23)
(4,-23)
(4,11)
};
\node at (axis cs:5,-7) {\textcolor{green}{\large $\{\Set{4,5},\Set{1}$}};
\node at (axis cs:5,-15) {\textcolor{green}{\large $\Set{2},\Set{3} \}$}};

\addplot [
color=orange,
solid,
line width=1.5pt,
mark=asterisk,
mark options={solid}
]
table[row sep=crcr]{
4 -3\\
0 -23 \\
};

\addplot[area legend,solid,fill=orange!20,opacity=4.000000e-01]coordinates {
(4,11)
(4,-23)
(0,-23)
(0,11)
};
\node at (axis cs:2,-2) {\textcolor{orange}{\large $\Set{\Set{1},\dotsc,\Set{5}}$}};

\end{axis}
\end{tikzpicture}%
}}
	\caption{The piecewise linear increasing Dilworth truncation $\FuHat{\alpha}(V_i)$ in $\alpha$ and the segmented partition $\Qat{\alpha}{V_i}$ obtained at the end of each iteration $i$ of the PAR Algorithm when it is applied to the $5$-user system in Example~\ref{ex:main}.}
	\label{fig:PSP}
\end{figure*}
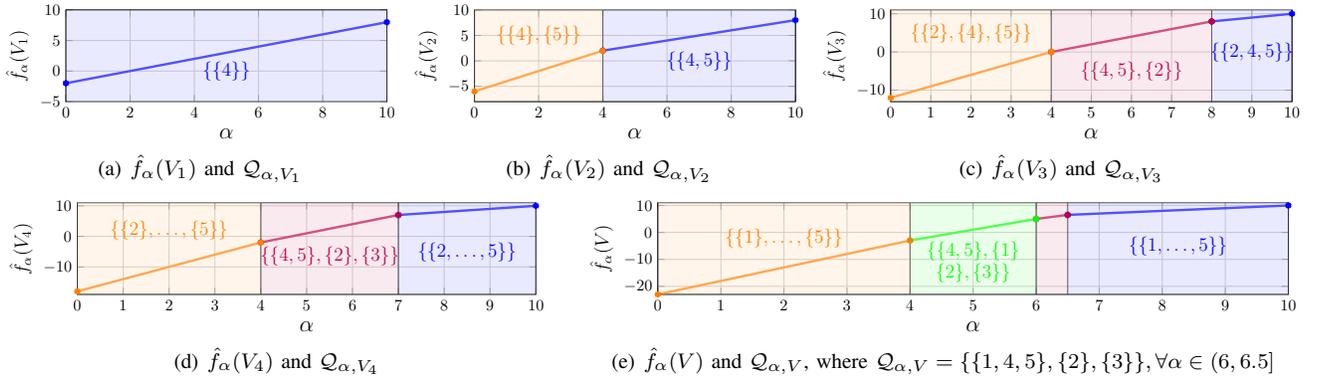

\begin{example} \label{ex:ParAlgo}
    We apply the PAR algorithm to the system in Example~\ref{ex:main}. First, initiate $r_{\alpha,i} = \alpha - H(V) = \alpha - 10$ for all $i \in V$ and $\alpha$. For $i=1$, we get $\Qat{\alpha}{V_1} = \Set{\Set{4}}$ and $r_{\alpha,4}=\Fu{\alpha}(\Set{4}) = \alpha - 2$ for all $\alpha$. See Fig.~\ref{fig:PSP}(a).
    As shown in Example~\ref{ex:PP}, for $i = 2$, we get $\TU{\alpha}{V_2}$ in \eqref{eq:ExU} so that the updated $\rv_{\alpha,V_2}$ and $\Qat{\alpha}{V_2}$ are in \eqref{eq:ExQatR}; for $i = 3$, we get $\TU{\alpha}{V_3}$ in \eqref{eq:UpV3} so that the updated $\rv_{\alpha,V_3}$ and $\Qat{\alpha}{V_3}$ are in \eqref{eq:UpdateV3}. See Fig.~\ref{fig:PSP}(b) and (c), respectively.

    For $i = 4$ and $\phi_4 = 3$, consider the problem $\min\Set{ \FuU{\alpha}(\TX) \colon \Set{3} \in \X \subseteq \Qat{\alpha}{V_4} }$ where $\Qat{\alpha}{V_4} = \Qat{\alpha}{V_3} \sqcup \Set{\Set{3}}$. We have the critical points $\alpha_1 = 7$ and $\alpha_0 = H(V) = 10$ with $\TUp{1} = \Set{3}$ and $\TUp{0} = \Set{2,3,4,5}$ such that
    $ \Set{3} = \TUp{1} \subsetneq  \TUp{0} = V_4 $
    and
    \begin{equation}
        \TU{\alpha}{V_4} = \begin{cases}
                             \Set{3} & \alpha \in [0,7],\\
                             \Set{2,3,4,5} & \alpha \in (7,10].
                         \end{cases} \nonumber
    \end{equation}
    We use $\TU{\alpha}{V_4}$ to update $\rv_{\alpha,V}$ and $\Qat{\alpha}{V_4}$ for all $\alpha$ as in step~\ref{step:UpdatesPar} and get
    \begin{equation} \label{eq:UpV4}
        \begin{aligned}
            & \rv_{\alpha,V} = \begin{cases}
                                    (\alpha-10, \alpha-6, \alpha-6, \alpha-2, \alpha-4) & \alpha \in [0,4], \\
                                    (\alpha-10, \alpha-6, \alpha-6, \alpha-2, 0) & \alpha \in (4,7], \\
                                    (\alpha-10, \alpha-6, 8-\alpha, \alpha-2,0) & \alpha \in (7,8], \\
                                    (\alpha-10, 2, 0, \alpha-2, 0)  & \alpha \in (8,10],
                                \end{cases}\\
            & \Qat{\alpha}{V_4} = \begin{cases}
                                    \Set{\Set{2},\Set{3},\Set{4},\Set{5}} & \alpha \in [0,4], \\
                                    \Set{\Set{4,5},\Set{2},\Set{3}} & \alpha \in (4,7], \\
                                    \Set{\Set{2,3,4,5}} & \alpha \in (7,10].
                                \end{cases}
        \end{aligned}
    \end{equation}
    See Fig.~\ref{fig:PSP}(d).

    For $i = 5$ and $\phi_5 = 1$, we have the critical points for the problem $\min\Set{ \FuU{\alpha}(\TX) \colon \Set{1} \in \X \subseteq \Qat{\alpha}{V} }$ being $\alpha_2 = 6$, $\alpha_1 = 6.5$ and $\alpha_0 = H(V) = 10$ with $\TUp{2} = \Set{1}$, $\TUp{1} = \Set{1,4,5}$ and $\TUp{0} = \Set{1,\dotsc,5}$ such that
    \begin{equation} \label{eq:UpV5U}
        \TU{\alpha}{V} = \begin{cases}
                             \Set{1} & \alpha \in [0,6],\\
                             \Set{1,4,5} & \alpha \in (6,6.5], \\
                             \Set{1,\dotsc,5} & \alpha \in (6.5,10].
                         \end{cases}
    \end{equation}
    After the updates in step~\ref{step:UpdatesPar},
    \begin{equation} \label{eq:UpV5}
        \begin{aligned}
            & \rv_{\alpha,V} = \begin{cases}
                                    (\alpha-5, \alpha-6, \alpha-6, \alpha-2, \alpha-4) & \alpha \in [0,4] ,\\
                                    (\alpha-5, \alpha-6, \alpha-6, \alpha-2, 0) & \alpha \in (4,6] ,\\
                                    (1, \alpha-6, \alpha-6, \alpha-2, 0) & \alpha \in (6,6.5] ,\\
                                    (14-2\alpha, \alpha-6, \alpha-6, \alpha-2, 0) & \alpha \in (6.5,7] ,\\
                                    (0, \alpha-6, 8-\alpha, \alpha-2, 0) & \alpha \in (7,8] ,\\
                                    (0, 2, 0, \alpha-2, 0) & \alpha \in (8,10] ,
                                \end{cases} \\
            & \Qat{\alpha}{V} = \begin{cases}
                                    \Set{\Set{1},\dotsc,\Set{5}} & \alpha \in [0,4], \\
                                    \Set{\Set{4,5},\Set{1},\Set{2},\Set{3}} & \alpha \in (4,6], \\
                                    \Set{\Set{1,4,5},\Set{2},\Set{3}} & \alpha \in (6,6.5], \\
                                    \Set{\Set{1,\dotsc,5}} & \alpha \in (6.5,10].
                                \end{cases}
        \end{aligned}
    \end{equation}
    See Fig.~\ref{fig:PSP}(e). For the final segmented partition $\Qat{\alpha}{V}$, the corresponding PSP has the critical points $\alphap{3} = 4$, $\alphap{2} = 6$ and $\alphap{1} = 6.5$ and $\alphap{0} = H(V) = 10$ with $\Patp{3} = \Set{\Set{1},\dotsc,\Set{5}}$, $\Patp{2} = \Set{\Set{4,5},\Set{1},\Set{2},\Set{3}}$, $\Patp{1} = \Set{\Set{1,4,5},\Set{2},\Set{3}}$ and $\Patp{0} = \Set{\Set{1,\dotsc,5}}$ so that we know $\RACO(V) = \alphap{1} = 6.5$ is the minimum sum-rate for the asymptotic model and $\Patp{1} = \Set{\Set{4,5,1},\Set{2},\Set{3}}$ is fundamental partition. We also know an optimal achievable rate vector $\rv_{6.5,V} = (1,0.5,0.5,4.5,0) \in \RRACO(V)$, which has the LCM $|\Patp{1}| - 1 = 2$ so that it is implementable by network coding schemes with $2$-packet-splitting in CCDE.
    The results also provide the solution to the non-asymptotic model: the minimum sum-rate is $\RNCO(V) = \lceil \RACO(V) \rceil = 7$ and $\rv_{7,V} = (0,1,1,5,0) \in \RRNCO(V)$ is an optimal achievable rate vector.\footnote{We used the same source model in Example~\ref{ex:main} as in \cite{Ding2018IT} so that the results can be compared and verified: the minimum sum-rate solutions to both asymptotic and non-asymptotic model are consistent with \cite{Ding2018IT}. }
\end{example}

The remaining problem is how to obtain $\alpha_j$'s and $\TUp{j}$'s in Lemma~\ref{lemma:PP}.
Recall that, $\alpha_j$'s and $\TUp{j}$'s are used to update $\Qat{\alpha}{V_i}$ for all $\alpha$ in step~\ref{step:Updates} of Algorithm~\ref{algo:ParAlgo} and the resulting $\Qat{\alpha}{V_i}$ determines the PSP of $V_i$.
We can still use Lemma~\ref{lemma:AlphaAdapt} in Appendix~\ref{app:AlphaAdapt} to adapt the value of $\alpha$ to search for $\TUp{j}$. This results in the $\StrMap$ algorithm in Algorithm~\ref{algo:StrMap}:
All $\TUp{j}$'s are determined by the call $\StrMap(\Set{\Set{m} \colon m \in V_i},\Set{V_i})$. The corresponding critical values $\alpha_j$ can be obtained by the property of the strict strong map below.

\begin{lemma}[{\cite[Theorem~31]{Fujishige2009PP}}] \label{lemma:CrVals}
    For all $\alpha_j$'s and $\TUp{j}$'s that characterize $\TU{\alpha}{V_i}$ of the minimal minimizer of $\min\Set{ \FuU{\alpha}(\TX) \colon \Set{\phi_i} \in \X \subseteq \Qat{\alpha}{V_i}}$ in Lemma~\ref{lemma:PP},\footnote{The rate vector $\rv_{\alpha,V}$ remains piecewise linear in $\alpha$ in Algorithm~\ref{algo:ParAlgo} so that $r_{\alpha_j} ( \TUp{j-1} \setminus \TUp{j})$ is a segmented linear function of $\alpha$. So, $\alpha_j$ can be determined by solving the linear equation $r_{\alpha} ( \TUp{j-1} \setminus \TUp{j}) = H(\TUp{j-1}) - H(\TUp{j})$.}
    $$ r_{\alpha_j} ( \TUp{j-1} \setminus \TUp{j}) = H(\TUp{j-1}) - H(\TUp{j}), \ \forall j \in \Set{1,\dotsc,q}. \IEEEQEDhereeqn $$
\end{lemma}

      \begin{algorithm} [t]
	       \label{algo:StrMap}
	       \small
	       \SetAlgoLined
	       \SetKwInOut{Input}{input}\SetKwInOut{Output}{output}
	       \SetKwFor{For}{for}{do}{endfor}
            \SetKwRepeat{Repeat}{repeat}{until}
            \SetKwIF{If}{ElseIf}{Else}{if}{then}{else if}{else}{endif}
	       \BlankLine
           \Input{$\Pat_d,\Pat_u \in \Pi(V_i)$ such that $\Pat_d \prec \Pat_u$ (We assume the $\Qat{\alpha}{V_i}$ and $\FuU{\alpha}$ for all $\alpha$ are the global variables.)}
	       \Output{A subset of $\Set{\TUp{j} \colon j \in \Set{0,\dotsc,q}}$ for the problem $\min \Set{ \FuU{\alpha}(\TX) \colon \Set{\phi_i} \in \X \subseteq \Qat{\alpha}{V_i}}$ in step~\ref{step:PP} of Algorithm~\ref{algo:ParAlgo}.}
	       \BlankLine
            $\alpha \coloneqq H(V) - \frac{ H[\Pat_d] - H[\Pat_u] }{ |\Pat_d| - |\Pat_u| }$\;
            $\SU \coloneqq \bigcap \argmin \Set{ \FuU{\alpha}(\TX) \colon \Set{\phi_i} \in \X \subseteq \Qat{\alpha}{V_i}} $\;
            $\Pat \coloneqq (\Qat{\alpha}{V_i} \setminus \SU) \sqcup \Set{ \TSU }$\;
            \lIf{$\Pat = \Pat_d$}{\Return $\Set{\TSU}$}
            \lElse{ \Return $\StrMap(\Pat_d,\Pat) \cup \StrMap(\Pat,\Pat_u)$ }
	   \caption{Strong Map (StrMap) Algorithm}
	   \end{algorithm}

\subsection{Minimum Weighted Sum-rate Problem}
\label{sec:WeightedSum}

Let $\wv_V = (w_i \colon i \in V) \in \RealPP^{|V|}$ be a positive weight vector and $\wv_V^{\intercal} \rv_V = \sum_{i \in V} w_i r_i$ be the weighted sum-rate of $\rv_V$. The minimum weighted sum-rate problem is to search a rate vector that minimizes the $\wv_V^{\intercal} \rv_V$ in the optimal rate region:
\begin{subequations} \label{eq:WeightSum}
    \begin{align}
        &\min\Set{ \wv_V^{\intercal} \rv_V \colon \rv_V \in \RRACO(V)}, \label{eq:WeightSumACO} \\
        &\min\Set{ \wv_V^{\intercal} \rv_V \colon \rv_V \in \RRNCO(V)}, \label{eq:WeightSumNCO}
    \end{align}
\end{subequations}
for the asymptotic and non-asymptotic models, respectively.
It is shown in \cite{MiloIT2016,CourtIT2014} that, by choosing a proper linear ordering $\Phi$, the solution is returned by the $\CoordSatCapFus$ algorithm. This method is described in \cite[Theorem~35]{Ding2018IT} as follows. For a given weight vector $\wv_V$, we call $\Phi = (\phi_1,\dotsc,\phi_{|V|})$ a linear ordering w.r.t. $\wv_V$ if $w_{\phi_1} \leq \dotsc \leq w_{\phi_{|V|}}$ and the calls $\CoordSatCapFus(\RACO(V), H, V, \Phi)$ and $\CoordSatCapFus(\RNCO(V), H, V, \Phi)$ return the minimizers of \eqref{eq:WeightSumACO} and \eqref{eq:WeightSumNCO}, respectively.
By knowing that that $\rv_{\alpha,V}$ returned by the PAR algorithm is exactly the same as the one returned by the call $\CoordSatCapFus(\alpha,H,V,\Phi)$ for all $\alpha$ and the relations $B(\FuHat{\RACO(V)}) = \RRACO(V)$ and $B(\FuHat{\RNCO(V)}) \cap \Z^{|V|} = \RRNCO(V)$, the following properties hold straightforwardly.\footnote{In fact, for any achievable sum-rate $\alpha \geq \RACO(V)$ and a linear ordering $\Phi$ w.r.t. a given $\wv_V$, the call $\CoordSatCapFus(\alpha,H,V,\Phi)$ returns a $\rv_{\alpha,V}$ that minimizes $\min\Set{ \wv_V^{\intercal} \rv_V \colon \rv_V \in \RRCO(V), r(V) = \alpha}$. }

\begin{corollary} \label{coro:MinWeightedSum}
    For a given weight vector $\wv_V$, choose the linear ordering $\Phi$ w.r.t. $\wv_V$. The rate vector $\rv_{\alpha,V}$ returned by the call $\PAR(H,V,\Phi)$ provides solutions to the minimum weighted sum-rate problems \eqref{eq:WeightSum} at $\alpha = \RACO(V)$ and $\alpha = \RNCO(V)$, i.e., $\rv_{\RACO(V),V} \in \argmin\Set{ \wv_V^{\intercal} \rv_V \colon \rv_V \in \RRACO(V)}$ and $\rv_{\RNCO(V),V} \in \argmin\Set{ \wv_V^{\intercal} \rv_V \colon \rv_V \in \RRNCO(V) }$, respectively. \hfill\IEEEQED
\end{corollary}


\subsection{Complexity}
\label{sec:Complexity}

The PSP invokes $|V|$ calls of the $\StrMap$ algorithm. As explained in Appendix~\ref{app:PSFM}, the $\StrMap$ algorithm can be implemented by the PSFM algorithms in \cite{Fleischer2003PSFM,Nagano2007PSFM,IwataPSFM1997} that have the same asymptotic\footnote{Here, `asymptotic' refers to the asymptotic limits of the complexity notation $O(\cdot)$: for the actual running time $a(|V|)$, the asymptotic complexity is $O(b(|V|))$ if $\lim_{|V| \rightarrow \infty} \frac{a(|V|)}{b(|V|)} = c$ for some constant $c$.}
complexity as the SFM algorithm. Therefore, the minimum sum-rate problem in \eqref{eq:MinSumRatePat}, as well as the minimum weighted sum-rate problem in \eqref{eq:WeightSum}, for both asymptotic and non-asymptotic models can be solved by the $\PAR$ algorithm in $O(|V| \cdot \SFM(|V|))$ time. As compared to the existing computation time $O(|V|^2 \cdot \SFM(|V|))$ of the MDA algorithm in \cite{Ding2018IT} and the algorithms in \cite{CourtIT2014,MiloIT2016} for the finite linear source model, the complexity is reduced by a factor of $|V|$.
In addition, the $\PAR$ algorithm allows distributed computation. See Section~\ref{sec:Distr}.

\section{Related Problems}
\label{sec:Relation}

CO was first formulated in \cite{Csiszar2004} based on the secret key agreement problem for the purpose of determining the secret capacity $\SC(A)$, the largest rate that the secret key can be generated by the active users in $A \subseteq V$ with the rest users in $V \setminus A$ being the helpers. The secret capacity is shown in \cite[Example~4]{Csiszar2004} to be upper bounded by a multivariate mutual dependence, which is proved to be tight when $A = V$ in \cite{Chan2008tight}. The relationship with the PSP became clearer in the further studies on the case $A = V$ of the secret key agreement and the CO problems in \cite{ChanMMI,Ding2018IT}.
In this section, we show the contribution of the $\PAR$ algorithm to the existing related problems.

\subsection{Secret Capacity}

The secret capacity in the case when $A = V$ is \cite[Example 4]{Csiszar2004}\cite{Chan2008tight}\footnote{It is shown $\SC(V) \leq I(V)$ in \cite[Example 4]{Csiszar2004}, which is proved to be tight in \cite{Chan2008tight}}
\begin{equation}\label{eq:SC}
    \SC(V) = I(V),
\end{equation}
where
\begin{equation} \label{eq:MMI}
    \begin{aligned}
        I(V) & = \min_{\Pat \in \Pi(V) \colon |\Pat| > 1} \frac{D( P_{\RZ{V}} \| \prod_{C \in \Pat} P_{\RZ{C}} )}{|\Pat|-1} \\
             & = \min_{\Pat \in \Pi(V) \colon |\Pat| > 1} \frac{H[\Pat] - H(V)}{|\Pat|-1}.
    \end{aligned}
\end{equation}
The term $I(V)$ is called the \textit{shared information} in \cite{Prakash2016} and \emph{multivariate mutual information} in \cite{ChanMMI}, that measures the mutual dependence in $\RZ{V}$. Based on \eqref{eq:SC}, we have the duality relationship between $\SC(V)$ and $\RACO(V)$ \cite[Theorem 1]{Csiszar2004}\cite{ChanSuccessive,ChanSuccessiveIT,ChanMMI}:
\begin{equation} \label{eq:Dual}
    \begin{aligned}
        \RACO(V) & = H(V) - \SC(V) \\
                 & = H(V) - I(V),
    \end{aligned}
\end{equation}
which states that the omniscience is attained by the minimum sum-rate $\RACO(V)$ if the users in $V$ only exchange over broadcast channels the amount of information that is not known to all.
It is shown in \cite{ChanMMI} that $I(V) = H(V) - \alphap{1}$ and the fundamental partition $\Patp{1}$ is the finest/minimal minimizer of \eqref{eq:MMI}. Thus, the solutions to both the CO and the secret agreement problems are provided by the PSP.
The $\PAR$ algorithm reduces the existing complexity \cite{ChanMMI,Ding2016NetCod,Ding2018IT} for determining the secret capacity $\SC(V)$ from $O(|V|^2 \cdot \SFM(|V|))$ to $O(|V| \cdot \SFM(|V|))$.

\subsection{Clustering}
\label{sec:Clustering}


For the data points in $V$, let $f$ be the normalized\footnote{A set function $f$ is \emph{normalized} if $f(\emptyset) = 0$.}
submodular set function such that $f(X)$ measures the inhomogeneity of the data points in subset $X \subseteq V$.
The entropy function $H$ and (graph) cut function $\kappa$ are two typical examples of the inhomogeneity measures.
For a (non-overlapping) clustering result $\Pat$ such that $|\Pat| > \beta$ for some $\beta \in [0, |V|)$, the clustering cost $f[\Pat] = \sum_{C \in \Pat} f(C)$ is normalized by the increment number of clusters $|\Pat| - \beta$ and the problem $\min_{\Pat \in \Pat \colon |\Pat| > 1}\frac{f[\Pat]}{|\Pat| - \beta}$ is called the $\beta$-minimum average cost ($\beta$-MAC) clustering \cite{MinAveCost}, to which the solution for all $\beta$ is fully determined by the minimizers of the Dilworth truncation $\FuHat{\lambda}(V) = \min_{\Pat \in \Pi(V)} \Fu{\lambda}[\Pat] = \min_{\Pat \in \Pi(V)} \sum_{C \in \Pat} \Fu{\lambda}(C)$, where $\Fu{\lambda} = f(X) - \lambda$ for all $X \subseteq V$.
As stated in \cite[Lemma~3]{MinAveCost}, the problem of $\beta$-MAC clustering is equivalent to determining the PSP of $V$.

By letting $\lambda = f(V) - \alpha$, $\Fu{\lambda}$ is equivalent to $\Fu{\alpha}$ defined in Section~\ref{subsec:ExResults} and all results derived in this paper also hold for all $\lambda$.
The minimal minimizer $\Qat{\lambda}{V} = \bigwedge \argmin_{\Pat \in \Pi{V}} \Fu{\lambda}[\Pat] = \Qat{\alpha}{V}$ with the critical points $\lambdap{j} = H(V) - \alphap{j}$ for all $j \in \Set{0,\dotsc,p}$ determining the PSP of $V$. See also
Lemma~\ref{lemma:AlphaAdapt} in Appendix~\ref{app:AlphaAdapt}.
The $1$-MAC clustering (when $\beta = 1$) is of particular interest in that it generalizes the network strength and can be extended to an information-theoretic clustering framework.

\subsubsection{Network Strength and Pairwise Independent Network (PIN) Model}
\label{subsec:PIN}

For $f$ being the cut function $\kappa$ of a graph, the $1$-MAC clustering determines the network strength $\sigma(V)$ \cite{Cunningham1985NetStrength}:
\begin{equation}\label{eq:NetStrength}
    \sigma(V) = \frac{1}{2} \min_{\Pat \in \Pi(V) \colon |\Pat| > 1} \frac{\Cut[\Pat]}{|\Pat|-1},
\end{equation}
where each $\Pat \in \Pi(V)$ denotes a multi-way cut. The cost this multi-way cut incurs is $\Cut[\Pat] = \sum_{C \in \Pat} \Cut(C)$. The network strength is a measure of connectivity in the view of the optimal network attack problem \cite{Cunningham1985NetStrength}\footnote{For an attacker, each partition $\Pat$ is regarded as a network decomposition method such that losses incurred by removing all edges connecting different subsets $C,C' \in \Pat$ and utility gained for the increased number of subgraphs $|\Pat| - 1$, the network strength $\sigma(V)$ is the largest per-subgraph payoff $\lambda$ such that the graph still remains intact \cite{Cunningham1985NetStrength}.}, which is determined by the first critical point in the PSP
$\sigma(V) = \frac{1}{2} \lambdap{1} = \frac{1}{2} (H(V) - \alphap{1})$.
It was shown in \cite{Cunningham1985NetStrength} that $\sigma(V)$ can be obtained in $O(|V|^2 \cdot \MaxFlow(|V|))$ time, where $\MaxFlow(|V|)$ denotes the complexity of the max-flow/min-cut algorithm \cite{MaxFlow1988} applied to a graph with $|V|$ nodes. This complexity is reduced to $O(|V| \cdot \MaxFlow(|V|))$ in \cite{Chen1994NetStrenght}.
The network strength also denotes the secret capacity in the pairwise independent network (PIN) source model \cite{GSK2007PIN,PIN2010,PINSteinerTree2010}, which has a graphical representation.

In a PIN model, we have each terminal being $\RZ{i} = (\RW{ii'} \colon i' \in V \setminus \Set{i})$, where the pairs of r.v.s in $\Set{(\RW{ii'},\RW{i'i}) \colon i,i' \in V, i \neq i'}$ are independent of each other.
The entropy function is $H(V) = \sum_{i,i' \in V \colon i \neq i'} H(\RW{ii'},\RW{i'i})$ and $H(X) = \sum_{i,i' \in X \colon i \neq i'} H(\RW{ii'},\RW{i'i}) + \sum_{i \in X, i' \notin X} H(\RW{ii'})$ for all $X \subseteq V$. Thus, the secret capacity~\eqref{eq:SC} reduces to \cite[Theorem~3.4]{PIN2010}
\begin{equation}\label{eq:SCPIN}
    \SC(V) = \sigma(V) = \frac{1}{2} \min_{\Pat \in \Pi(V) \colon |\Pat| > 1} \frac{\Cut[\Pat]}{|\Pat|-1},
\end{equation}
where $\Cut$ is the cut function of the undirected graph $G = (V,E)$ with the weight of each edge $(i,i')$ being $I(\RW{ii'};\RW{i'i})$. See Example~\ref{ex:AUX}.
The idea of solving the secret agreement problem in the PIN model via the tree packing algorithms, e.g., \cite{Kruskal1960},
is based on the relationship \cite[Section~5.1]{IBMRep2011}\cite{Tutte1961,Williams1961}: the maximum number of edge-disjoint spanning trees is $\lfloor \sigma(V) \rfloor$.
More directly, any algorithm determining the network strength $\sigma(V)$, e.g., \cite{Cunningham1985NetStrength,Chen1994NetStrenght}, can also be applied to the secret agreement problem in the PIN model.

\subsubsection{Information-theoretic Clustering}
\label{subsec:Clustering}

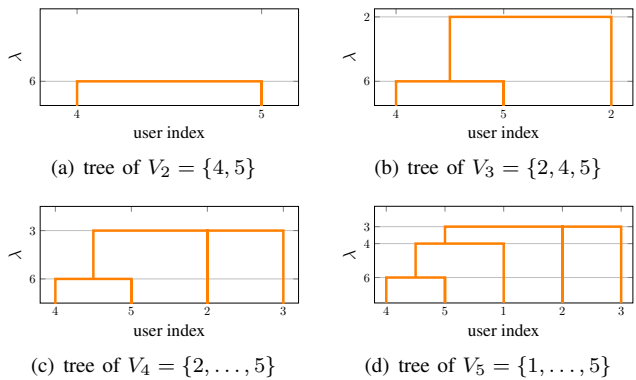
\begin{figure}[t]
	\centering
    \subfigure[tree of $V_2 = \Set{4,5}$]{\scalebox{0.47}{\begin{tikzpicture}

\begin{axis}[
width=3.5in,
height=1.7in,
xlabel= {\Large user index},
xmin=0.8,
xmax=2.2,
xtick={1,2},
xticklabels={$4$,$5$},
ylabel = {\Large $\lambda$},
ymin=0.5,
ymax=6.5,
ytick={2},
yticklabels={$6$},
ymajorgrids,
]

\addplot[color=orange, line width = 2] coordinates {
(1,0)
(1,2)
(2,2)
(2,0)
(2,2)
};

\end{axis}

\end{tikzpicture}}} \quad
    \subfigure[tree of $V_3 = \Set{2,4,5}$]{\scalebox{0.47}{\begin{tikzpicture}

\begin{axis}[
width=3.5in,
height=1.7in,
xlabel= {\Large user index},
xmin=0.8,
xmax=3.2,
xtick={1,2,3},
xticklabels={$4$,$5$,$2$},
ylabel = {\Large $\lambda$},
ymin=0.5,
ymax=6.5,
ytick={2,6},
yticklabels={$6$,$2$},
ymajorgrids,
]

\addplot[color=orange, line width = 2] coordinates {
(1,0)
(1,2)
(2,2)
(2,0)
(2,2)
};
\addplot[color=orange, line width = 2] coordinates {
(1.5,2)
(1.5,6)
(3,6)
(3,0)
};

\end{axis}

\end{tikzpicture}}} \\
    \subfigure[tree of $V_4 = \Set{2,\dotsc,5}$]{\scalebox{0.47}{\begin{tikzpicture}

\begin{axis}[
width=3.5in,
height=1.7in,
xlabel= {\Large user index},
xmin=0.8,
xmax=4.2,
xtick={1,2,3,4},
xticklabels={$4$,$5$,$2$,$3$},
ylabel = {\Large $\lambda$},
ymin=0.5,
ymax=6.5,
ytick={2,5},
yticklabels={$6$,$3$},
ymajorgrids,
]

\addplot[color=orange, line width = 2] coordinates {
(1,0)
(1,2)
(2,2)
(2,0)
(2,2)
};
\addplot[color=orange, line width = 2] coordinates {
(1.5,2)
(1.5,5)
(3,5)
(3,0)
(3,5)
(4,5)
(4,0)
};

\end{axis}

\end{tikzpicture}}} \quad
    \subfigure[tree of $V_5 = \Set{1,\dotsc,5}$]{\scalebox{0.47}{\begin{tikzpicture}

\begin{axis}[
width=3.5in,
height=1.7in,
xlabel= {\Large user index},
xmin=0.8,
xmax=5.2,
xtick={1,2,3,4,5},
xticklabels={$4$,$5$,$1$,$2$,$3$},
ylabel = {\Large $\lambda$},
ymin=0.5,
ymax=6.2,
ytick={2,4,5},
yticklabels={$6$,$4$,$3$},
ymajorgrids,
]

\addplot[color=orange, line width = 2] coordinates {
(1,0)
(1,2)
(2,2)
(2,0)
(2,2)
};
\addplot[color=orange, line width = 2] coordinates {
(1.5,2)
(1.5,4)
(3,4)
(3,0)
};

\addplot[color=orange, line width = 2] coordinates {
(2,4)
(2,5)
(4,5)
(4,0)
(4,5)
(5,5)
(5,0)
};

\end{axis}

\end{tikzpicture}}}
	\caption{For Fig.~\ref{fig:PSP}(b) to (e), replacing the horizontal axis $\alpha$ by $H(V) - \lambda$, we have the dendrogram in (a) to (d), respectively, each of which is the hierarchical clustering result of $V_i$ by the shared/multivariate mutual information measure $I(X)$ in \eqref{eq:MMI} (see also Section~\ref{subsec:Clustering}) \cite{Chan2016InfoClustering}. These dendrograms can also be obtained at the end of each iteration of the $\DistrPAR$ algorithm (Algorithm~\ref{algo:DistrParAlgo}).}
	\label{fig:TreeMain}
\end{figure}

The authors in \cite{Chan2016InfoClustering} extended the $1$-MAC ($\beta = 1$) clustering problem based on the measure $I(V)$ in \eqref{eq:MMI}. It is shown in \cite[Theorem~3]{Chan2016InfoClustering} that $C = \bigcup \Set{X \subseteq V \colon I(X) > \lambda, |X| > 1}$ for all $C \in \Qat{\lambda}{V}$ such that $|C| > 1$.
The interpretation is that any nonsingleton $C \in \Qat{\lambda}{V}$ is the maximal subset with similarity $I(C)$ strictly greater than a given threshold $\lambda$. In this sense, all critical points $\lambdap{j}$ and partitions $\Patp{j}$'s in the PSP form a hierarchical clustering result.
For example, replacing $\alpha$ by $H(V) - \lambda$ for $\Qat{\alpha}{V_4}$ in \eqref{eq:UpV4}, we have
\begin{equation} \label{eq:UpV4Lambda}
    \Qat{\lambda}{V_4} = \begin{cases}
                                    \Set{\Set{2,3,4,5}} & \lambda \in [0,3), \\
                                    \Set{\Set{4,5},\Set{2},\Set{3}} & \lambda \in [3,6), \\
                                    \Set{\Set{2},\Set{3},\Set{4},\Set{5}} & \lambda \in [6,+\infty).
                                \end{cases}
\end{equation}
corresponding to the dendrogram in Fig.~\ref{fig:TreeMain}(c).
Here, $I(\Set{4,5}) = 6$ and $I(\Set{4,5} \sqcup X) \leq \lambda$ for all $X \subseteq \Set{2,3}$ and any similarity threshold $\lambda \in [3,6)$, i.e., $\Set{4,5}$ is the maximal subset with a shared information $I(\Set{4,5})$ strictly greater than $\lambda$. Therefore, $\Set{4,5} \in \Qat{\lambda}{V_4}$ for all $\lambda \in [3,6)$.
In the region $\lambda \in [0,3)$, $I(\Set{4,5}) > \lambda$ means that users $4$ and $5$ should be clustered, i.e., $\Set{4,5}$ must be contained in some cluster/subset in $\Qat{\lambda}{V_4}$. But, in this case, $\Set{2,\dotsc,5}$ is the maximal subset with $I(\Set{2,\dotsc,5}) > \lambda$.
%
%
In Part II of this paper \cite{DingITSO2019}, we show that the dendrogram indicates a bottom-up successive omniscience approach for the asymptotic model.

The $\PAR$ algorithm solves the $\beta$-MAC and information-theoretic clustering problems in $O(|V| \cdot \SFM(|V|))$ time, which is faster than the existing computation time $O(|V|^2 \cdot \SFM(|V|))$ \cite[Algorithm~SPLIT]{MinAveCost} \cite[Algorithm~3]{Chan2016InfoClustering}.
In \eqref{eq:MMI}, replacing the entropy function $H$ by the cut function $\Cut$ with $\Cut(V) = 0$ of a graph, we have the network strength $\sigma(V) = \frac{1}{2} I(V)$. 
The call $\PAR(\Cut, V, \Phi)$ for any linear ordering $\Phi$ returns the PSP of the graph, of which the first critical value determines to the network strength in \eqref{eq:NetStrength} and the secret capacity $\SC(V)$ of the PIN model in \eqref{eq:SCPIN}: $\SC(V) = \sigma(V) = \frac{1}{2}\lambdap{1}$. In this call, the solution to the $\SFM$ problem \eqref{eq:Fusion} is the min-cut and the $\StrMap$ algorithm (Algorithm~\ref{algo:StrMap}) can be implemented by the parametric max-flow algorithm \cite{ParMaxFlow1989}.
The parametric max-flow algorithm \cite{ParMaxFlow1989} and max-flow algorithm \cite{MaxFlow1988} complete in the same time. The complexity of $\PAR$ in this case is $O(|V| \cdot \MaxFlow(|V|))$.\footnote{For the graph model, while the algorithm in \cite{Chen1994NetStrenght} only searches the first critical value, the $\PAR$ algorithm returns all critical values of the PSP. The PSP of the graph model also provides the solution to the optimal network attack problem \cite{Cunningham1985NetStrength} for all $\lambda$, where $\lambda$ is interpreted as the per-subgraph payoff.}

Independently, Kolmogorov proposed another parametric algorithm in \cite[Fig.~3]{Kolmogorov2010NetPSP} for determining the PSP specifically for the graph model, which also completes in $O(|V| \cdot \MaxFlow(|V|))$ time.
In Appendix~\ref{app:Kolmogorov}, we show that Kolmogorov's algorithm \cite[Fig.~3]{Kolmogorov2010NetPSP} is also based on a non-strict strong map property and propose a $\StrMapKomo$ algorithm to allow it to be applicable to general submodular functions $f$.

\section{Distributed Computation}
\label{sec:Distr}

An observation about Fig.~\ref{fig:PSP} is that the minimum sum-rate problem can be solved when the size of the ground set $V_i$ is gradually increasing in the order of $i = 1,2,\dotsc,|V|$.
Replacing $\alpha$ by $H(V) - \lambda$ in the $\PAR$ algorithm, we have $\Qat{\lambda}{V_i} = \bigwedge \argmin_{\Pat \in \Pi(V_i)} \Fu{\lambda}[\Pat]$ for all $\lambda$ at the end of each iteration $i$.
For $\lambda$ interpreted as the estimate of the shared/multivariate mutual information, or the secret capacity, based on the dual relationship \eqref{eq:Dual}, $\Fu{\lambda}(X) = H(X) - \lambda, \forall X \subseteq V$ is the called \emph{residual entropy} function and the critical points of $\Qat{\lambda}{V_i}$ are $\lambdap{j} = H(V) - \alphap{j}$ for all $j \in \Set{0,\dotsc,p}$.
To run the $i$th iteration, only the knowledge of the first $i$ users in $V_i = \Set{\phi_1,\dotsc,\phi_i}$ is required: the value of $H(X)$ for all $X \subseteq V_i$ and the rate vector $\rv_{\alpha,V_{i-1}}$ that has been updated in the previous iterations. This suggests a distributed and adaptive computation of the PSP of $V$ by the $\PAR$ algorithm as in Algorithm~\ref{algo:DistrParAlgo}.\footnote{In Algorithm~\ref{algo:DistrParAlgo} `for all $\lambda$' means for all nonnegative values of $\lambda$ since the minimizer of $\min_{\Pat \in \Pi(V)} \Fu{\lambda}[\Pat]$ is always $\Set{V}$ for all $\lambda < 0$.}
For example, when the $\DistrPAR$ algorithm is applied to the $5$-user system in Example~\ref{ex:main}, we get the Dilworth truncation $\FuHat{\lambda}(V_i)$ being the same as in Fig.~\ref{fig:PSP} for $\lambda = H(V) - \alpha$.
We show another example below.

      \begin{algorithm} [t]
	       \label{algo:DistrParAlgo}
	       \small
	       \SetAlgoLined
	       \SetKwInOut{Input}{input}\SetKwInOut{Output}{output}
	       \SetKwFor{For}{for}{do}{endfor}
            \SetKwRepeat{Repeat}{repeat}{until}
            \SetKwIF{If}{ElseIf}{Else}{if}{then}{else if}{else}{endif}
	       \BlankLine
           \Input{$f$, $V$ and $\Phi$}
	       \Output{segmented variables $\rv_V \in B(\FuHat{\lambda})$ and $\Qat{\lambda}{V} = \bigwedge \argmin_{\Pat\in \Pi(V)} \Fu{\lambda}[\Pat]$ for all $\lambda$}
	       \BlankLine
            Let user $\phi_1$ initiate $ r_{\lambda,\phi_1} \coloneqq - \lambda$ and $\Qat{\alpha}{V_1} \coloneqq \Set{\Set{\phi_1}}$ for all $\lambda$ and pass to user $\phi_2$\;
            \For{$i=2$ \emph{\KwTo} $|V|$, let user $\phi_i$}{
                $r_{\lambda,\phi_i} \coloneqq -\lambda$ for all $\lambda $\;
                $\Qat{\lambda}{V_i} \coloneqq \Qat{\lambda}{V_{i-1}} \sqcup \Set{\Set{\phi_i}}$ for all $\lambda$\;
                For function $\FuU{\lambda}(\TX) = \Fu{\lambda}(\TX) - r_{\lambda}(\TX),\forall \X \subseteq \Qat{\lambda}{V_i}$, obtain the critical points $\Set{\lambda_j \colon j \in \Set{0,\dotsc,q}}$ and $\Set{\TUp{j} \colon j \in \Set{0,\dotsc,q}}$ that determine the minimal minimizer $\U{\lambda}{V_i}$ of $\min\Set{ \FuU{\lambda}(\TX) \colon \Set{\phi_i} \in \X \subseteq \Qat{\lambda}{V_i}}$ for all $\lambda$ by the $\StrMapDistPAR$ algorithm (Algorithm~\ref{algo:StrMapAlt})\;  \label{step:PPDistrPar}
                \lFor{$j=0$ \emph{\KwTo} $q$}{
                    $$ \rv_{\lambda,V} \coloneqq \rv_{\lambda,V} + \FuU{\lambda}(\TUp{j}) \chi_{\phi_i}; $$
                    $$ \Qat{\lambda}{V_i} \coloneqq (\Qat{\lambda}{V_i} \setminus \Up{j}) \sqcup \Set{ \TUp{j} }; $$
                    for all $\lambda \in [\lambda_j, \lambda_{j+1})$\label{step:UpatesDistrPar}}
                Pass the results $\rv_{\lambda,V_i}$ and $\Qat_{\lambda,V_i}$ as well as function $f(X)$ for all $X \subseteq V_i$ to user $\phi_{i+1}$\; \label{step:PPDistrForward}
            }
            \Return $\rv_V$ and $\Qat{\lambda}{V}$ for all $\lambda$\;
	   \caption{Distributed computation of PSP by $\PAR$ algorithm ($\DistrPAR$)}
	   \end{algorithm}

\begin{figure*}[t]
	\centering
    \subfigure[$\FuHat{\lambda}(V_1)$ and $\Qat{\lambda}{V_1}$]{\scalebox{0.6}{
%
%
\definecolor{mycolor1}{rgb}{1,0,1}%

\begin{tikzpicture}[
every pin/.style={fill=yellow!50!white,rectangle,rounded corners=3pt,font=\tiny},
every pin edge/.style={<-}]

\begin{axis}[%
width=2.8in,
height=0.8in,
scale only axis,
xmin=0,
xmax=2,
xlabel={\Large $\lambda$},
xmajorgrids,
ymin=-0.5,
ymax=2.5,
ylabel={\large $\FuHat{\lambda}(V_1)$},
ymajorgrids,
legend style={at={(0.65,0.05)},anchor=south west,draw=black,fill=white,legend cell align=left}
]

\addplot [
color=blue,
solid,
line width=1.5pt,
mark=asterisk,
mark options={solid}
]
table[row sep=crcr]{
0 2\\
3 -1\\
};

\addplot[area legend,solid,fill=blue!20,opacity=4.000000e-01]coordinates {
(2,3)
(2,-8)
(0,-8)
(0,3)
};
\node at (axis cs:0.8,0.6) {\textcolor{blue}{\large $\Set{\Set{1}}$}};


\end{axis}
\end{tikzpicture}%
}} \quad
    \subfigure[$\FuHat{\lambda}(V_2)$ and $\Qat{\lambda}{V_2}$]{\scalebox{0.6}{
%
%
\definecolor{mycolor1}{rgb}{1,0,1}%

\begin{tikzpicture}[
every pin/.style={fill=yellow!50!white,rectangle,rounded corners=3pt,font=\tiny},
every pin edge/.style={<-}]

\begin{axis}[%
width=2.8in,
height=0.8in,
scale only axis,
xmin=0,
xmax=3,
xlabel={\Large $\lambda$},
xmajorgrids,
ymin=-2.5,
ymax=3.5,
ylabel={\large $\FuHat{\lambda}(V_2)$},
ymajorgrids,
legend style={at={(0.65,0.05)},anchor=south west,draw=black,fill=white,legend cell align=left}
]

\addplot [
color=blue,
solid,
line width=1.5pt,
mark=asterisk,
mark options={solid}
]
table[row sep=crcr]{
0 3\\
1 2 \\
};

\addplot[area legend,solid,fill=blue!20,opacity=4.000000e-01]coordinates {
(0,10)
(0,-8)
(1,-8)
(1,10)
};
\node at (axis cs:0.5,1.2) {\textcolor{blue}{\large $\Set{\Set{1,2}}$}};

\addplot [
color=orange,
solid,
line width=1.5pt,
mark=asterisk,
mark options={solid}
]
table[row sep=crcr]{
1 2\\
4 -4\\
};

\addplot[area legend,solid,fill=orange!20,opacity=4.000000e-01]coordinates {
(1,10)
(1,-8)
(4,-8)
(4,10)
};
\node at (axis cs:2.5,1) {\textcolor{orange}{\large $\Set{\Set{1},\Set{2}}$}};

\end{axis}
\end{tikzpicture}%
}} \quad
    \subfigure[$\FuHat{\lambda}(V_3)$ and $\Qat{\lambda}{V_3}$]{\scalebox{0.6}{
%
%
\definecolor{mycolor1}{rgb}{1,0,1}%

\begin{tikzpicture}[
every pin/.style={fill=yellow!50!white,rectangle,rounded corners=3pt,font=\tiny},
every pin edge/.style={<-}]

\begin{axis}[%
width=2.8in,
height=0.8in,
scale only axis,
xmin=0,
xmax=3,
xlabel={\Large $\lambda$},
xmajorgrids,
ymin=-3.5,
ymax=3.5,
ylabel={\large $\FuHat{\lambda}(V_3)$},
ymajorgrids,
legend style={at={(0.65,0.05)},anchor=south west,draw=black,fill=white,legend cell align=left}
]

\addplot [
color=blue,
solid,
line width=1.5pt,
mark=asterisk,
mark options={solid}
]
table[row sep=crcr]{
0 3\\
1.5 1.5\\
};

\addplot[area legend,solid,fill=blue!20,opacity=4.000000e-01]coordinates {
(0,11)
(0,-13)
(1.5,-13)
(1.5,11)
};
\node at (axis cs:0.7,0) {\textcolor{blue}{\large $\Set{\Set{1,2,3}}$}};

\addplot [
color=purple,
solid,
line width=1.5pt,
mark=asterisk,
mark options={solid}
]
table[row sep=crcr]{
1.5 1.5\\
4 -6\\
};

\addplot[area legend,solid,fill=purple!20,opacity=4.000000e-01]coordinates {
(1.5,11)
(1.5,-13)
(4,-13)
(4,11)
};
\node at (axis cs:2.4,1.5) {\textcolor{purple}{\large $\Set{\Set{1},\Set{2},\Set{3}}$}};

\end{axis}
\end{tikzpicture}%
}} \\
    \subfigure[$\FuHat{\lambda}(V)$ and $\Qat{\lambda}{V}$]{\scalebox{0.6}{
%
%
\definecolor{mycolor1}{rgb}{1,0,1}%

\begin{tikzpicture}[
every pin/.style={fill=yellow!50!white,rectangle,rounded corners=3pt,font=\tiny},
every pin edge/.style={<-}]

\begin{axis}[%
width=4in,
height=0.8in,
scale only axis,
xmin=0,
xmax=4,
xlabel={\Large $\lambda$},
xmajorgrids,
ymin=-8.5,
ymax=4.5,
ylabel={\large $\FuHat{\lambda}(V)$},
ymajorgrids,
legend style={at={(0.65,0.05)},anchor=south west,draw=black,fill=white,legend cell align=left}
]

\addplot [
color=blue,
solid,
line width=1.5pt,
mark=asterisk,
mark options={solid}
]
table[row sep=crcr]{
0 4\\
1 3\\
};
\addplot[area legend,solid,fill=blue!20,opacity=4.000000e-01]coordinates {
(0,11)
(0,-19)
(1,-19)
(1,11)
};
\node at (axis cs:0.5,-2.5) {\textcolor{blue}{\large $\Set{\Set{1,\dotsc,4}}$}};

\addplot [
color=green,
solid,
line width=1.5pt,
mark=asterisk,
mark options={solid}
]
table[row sep=crcr]{
1 3\\
1.5 2\\
};

\addplot[area legend,solid,fill=green!20,opacity=4.000000e-01]coordinates {
(1,11)
(1,-19)
(1.5,-19)
(1.5,11)
};
\node at (axis cs:1.25,-0.5) {\textcolor{green}{\large $\{\{1,2,$}};
\node at (axis cs:1.25,-3.5) {\textcolor{green}{\large $3\},$}};
\node at (axis cs:1.25,-6.5) {\textcolor{green}{\large $\Set{4}\}$}};

\addplot [
color=orange,
solid,
line width=1.5pt,
mark=asterisk,
mark options={solid}
]
table[row sep=crcr]{
1.5 2\\
5 -12 \\
};

\addplot[area legend,solid,fill=orange!20,opacity=4.000000e-01]coordinates {
(1.5,11)
(1.5,-19)
(5,-19)
(5,11)
};
\node at (axis cs:3,0) {\textcolor{orange}{\large $\Set{\Set{1},\dotsc,\Set{4}}$}};

\end{axis}
\end{tikzpicture}%
}} \qquad \qquad \qquad
    \subfigure[PIN model $\RZ{V_3}$]{\scalebox{0.8}{\begin{tikzpicture}

\draw (-1.5,-1.5) circle (0.3);
\node at (-1.5,-1.5) {\Large $1$};

\draw (1.5,-1.5) circle (0.3);
\node at (1.5,-1.5) {\Large $2$};

\draw (0,0) circle (0.3);
\node at (0,0) {\Large $3$};

\draw (-1.3,-1.3) -- node [sloped,above] {$1$} (-0.25,-0.15);
\draw (0.25,-0.15) -- node [sloped,above] {$1$}(1.3,-1.3);
\draw (1.2,-1.5) -- node [above] {$1$}(-1.2,-1.5);

\end{tikzpicture} }}
	\caption{The piecewise linear decreasing Dilworth truncation $\FuHat{\lambda}(V_i)$ in $\lambda$ and the segmented partition $\Qat{\lambda}{V_i}$ obtained at the end of each iteration of the distributed PAR ($\DistrPAR$) algorithm (Algorithm~\ref{algo:DistrParAlgo}) when it is applied to the $4$-user system in Example~\ref{ex:AUX}. Here, $\Qat{\lambda}{V_i}$ in (a) to (d) characterizes the PSP of the system $V_i$. In addition, the first $3$-users in $V_3 = \Set{1,2,3}$ with $\RZ{\Set{1,2,3}}$ form a PIN model, for which, the secret agreement problem can be represented by the undirected graph in (e) with the weight of each edge $(i,i')$ being $I(\RZ{i};\RZ{i'})$. The strength of this graph is $\sigma(V_3) = 1.5$, which equals the secret capacity $\SC(V_3)$ and determines the minimum sum-rate $\RACO(V_3)$ based on the dual relationship \eqref{eq:Dual} $\RACO(V_3) = H(V_3) - \SC(V_3) = 1.5$. }
	\label{fig:PSPAux}
\end{figure*}
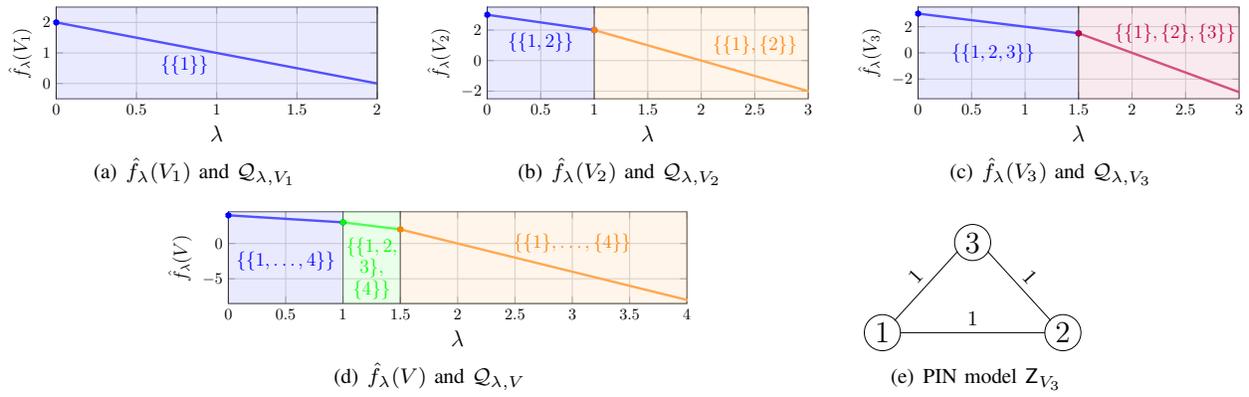

\begin{example}\label{ex:AUX}
       Consider a $4$-user system with
        \begin{equation}
            \begin{aligned}
                \RZ{1} & = (\RW{a},\RW{b}),  \qquad \RZ{2} & = (\RW{b},\RW{c}),   \\
                \RZ{3} & = (\RW{a},\RW{c}), \qquad \RZ{4} & = (\RW{c},\RW{d}),
            \end{aligned}  \nonumber
        \end{equation}
        where each $\RW{m}$ is an independent uniformly distributed random bit. Here, $\RZ{\Set{1,2,3}}$ forms a PIN model. The corresponding undirected graph $G = (V_3,E)$ is shown in Fig.~\ref{fig:PSPAux}(e), where the weight of edge $(i, i')$ is $I(\RZ{i},\RZ{i'})$.

        We run the $\DistrPAR$ algorithm in Algorithm~\ref{algo:DistrParAlgo} for $f = H$ and the linear ordering $\Phi = (1,2,3,4)$ as follows.
        First, user $1$ initiates $r_{\lambda,\phi_1} = r_{\lambda,1} = -\lambda$ and $\Qat{\alpha}{V_1} = \Set{\Set{1}}$ for all $\lambda \in [0,+\infty)$ and passes them to user $2$.
        User $2$ initiates $r_{\lambda,2} = - \lambda$ for all $\lambda$. For $\Qat{\lambda}{V_2} = \Qat{\lambda}{V_{1}} \sqcup \Set{\Set{2}} = \Set{\Set{1},\Set{2}}$, the minimal minimizer of $\min\Set{ \FuU{\lambda}(\TX) \colon \Set{2} \in \X \subseteq \Qat{\lambda}{V_2}}$ is $\U{\lambda}{V_2} = \Set{\Set{1},\Set{2}}$ for $\lambda \in [0,1)$ and $\U{\lambda}{V_2} = \Set{\Set{2}}$ for $\lambda \in [1,+\infty)$. After step~\ref{step:UpatesDistrPar}, he/she obtains $\Qat{\lambda}{V_2}$ as in Fig.~\ref{fig:PSPAux}(b) and the rate vector
        $$ \rv_{\lambda,V_2} = \begin{cases}
                                    (2-\lambda, 1) & \lambda \in [0,1), \\
                                    (2-\lambda,2-\lambda)  & \lambda \in [1,+\infty).
                                \end{cases}$$
        User $3$ obtains $\rv_{\lambda,V_2}$ and $\Qat{\lambda}{V_2}$ from user $2$ and gets
        $$ \U{\lambda}{V_3} = \begin{cases}
                                    \Set{\Set{1,2},\Set{3}} & \lambda \in [0,1), \\
                                    \Set{\Set{1},\Set{2},\Set{3}} & \lambda \in [1,1.5), \\
                                    \Set{\Set{3}} & \lambda \in [1.5,+\infty)
                            \end{cases} $$
        so that after step~\ref{step:UpatesDistrPar}, he/she obtains $\Qat{\lambda}{V_3}$ in Fig.~\ref{fig:PSPAux}(c) and
        $$ \rv_{\lambda,V_3} = \begin{cases}
                                    (2-\lambda,1,0) & \lambda \in [0,1), \\
                                    (2-\lambda,2-\lambda,\lambda-1) & \lambda \in [1,1.5), \\
                                    (2-\lambda,2-\lambda,2-\lambda)  & \lambda \in [1.5,+\infty),
                                \end{cases} $$
        Note, the first critical point $\lambdap{1} = 1.5$ equals the secret capacity $\SC(\Set{1,2,3})$ of the first three users.
        Alternatively, for $\kappa$ being the cut function of the undirected graph in Fig.~\ref{fig:PSPAux}(e), one can show that $\frac{1}{2} \min_{\Pat \in \Pi(V) \colon |\Pat| > 1} \frac{\Cut[\Pat]}{|\Pat|-1} = 1.5 = \lambdap{1} = \SC(\Set{1,2,3})$, which equals the network strength $\sigma(\Set{1,2,3})$.
        User $3$ passes $\Qat{\lambda}{V_3}$ and $\rv_{\lambda,V_3}$ to user $4$, where $\Qat{\lambda}{V}$ in Fig.~\ref{fig:PSPAux}(d) and
        \begin{equation} \label{eq:UpV4Aux}
            \rv_{\lambda,V} = \begin{cases}
                                        (2-\lambda,1,0,1) & \lambda \in [0,1), \\
                                        (2-\lambda,2-\lambda,\lambda-1,2-\lambda) & \lambda \in [1,1.5), \\
                                        (2-\lambda,2-\lambda,2-\lambda,2-\lambda)  & \lambda \in [1.5,+\infty)
                                    \end{cases}
        \end{equation}
        are obtained.

        For obtaining the minimal minimizer $\U{\lambda}{V}$ in each iteration $i$, one can still apply the $\StrMap$ algorithm (Algorithm~\ref{algo:StrMap}) to get $\Set{\TUp{j} \colon j \in \Set{0,\dotsc,q}}$. Based on Lemma~\ref{lemma:CrVals}, the critical points are $\lambda_j = H(V) - \alpha_j$. Independently, we propose a $\StrMapDistPAR$ algorithm in Appendix~\ref{app:PPAlt} that determines the segmented $\U{\lambda}{V_i}$.
\end{example}

In the $\DistrPAR$ algorithm, the complexity incurred at each user is $O(\SFM(|V|))$.
At the end of each iteration, $\Qat{\lambda}{V_i}$ determines all the critical points $\lambdap{j}$ and partitions $\Patp{j}$ in the PSP of $V_i$. The first critical point $\lambdap{1}$ equals the secret capacity $\SC(V_i)$ and shared/multivariate mutual information $I(V_i)$. For the CO problem in $V_i$, the value $\alphap{1} = H(V_i) - \lambdap{1}$ equals the minimum sum-rate $\RACO(V_i)$ and $\rv_{\lambdap{1},V_i}$ is an optimal rate vector in the asymptotic model; $\lceil \alphap{1} \rceil = H(V_i) - \lfloor \lambdap{1} \rfloor = \RNCO(V_i)$ and $\rv_{\lfloor \lambdap{1} \rfloor,V_i}$ is an optimal rate vector in the non-asymptotic model.
For example, consider the CO problem in the PIN model formed by the first 3 users in Example~\ref{ex:AUX}. The minimum sum-rate is $\RACO(V_3) = \alphap{1} = H(V_3) - \lambdap{1} = 1.5$ and $\rv_{1.5,V} = (0.5,0.5,0.5)$ is an optimal rate vector in the asymptotic model; The minimum sum-rate $\RNCO(V_3) = \lceil \alphap{1} \rceil = H(V_3) - \lfloor \lambdap{1} \rfloor = 2$ and $\rv_{1,V} = (1,1,0)$ is an optimal rate vector in the non-asymptotic model.
It means that the local omniscience problem in $V_i$ is solved before the global omniscience.
This fact will be utilized in Part II \cite{DingITSO2019} for solving the SO problem, where it is also shown that an optimal local omniscience in $X \subseteq V_i$ can be directly determined from $\rv_{\lambda,V_i}$.


$\DistrPAR$ is also an adaptive approach where $\Qat{\lambda}{V_i}$ is adapted from $\Qat{\lambda}{V_{i-1}}$ based on the minimal minimizer $\U{\lambda}{V_{i}}$ of $\min\Set{ \FuU{\lambda}(\TX) \colon \Set{\phi_i} \in \X \subseteq \Qat{\lambda}{V_{i-1}} \sqcup \Set{\phi_i}}$ for all $\lambda$. The value of $\Qat{\lambda}{V_i}$ converges to $\Qat{\lambda}{V}$ at the last user $\phi_{|V|}$.
This is particularly useful when the users complete recording their observations at different times. In this case, $\phi_i$ in the linear ordering $\Phi$ denotes that user $\phi_i$ is the $i$th user that finishes observing $\RZ{\phi_i}$.
Thus, instead of waiting for all users having the data ready, the $\PAR$ algorithm can be implemented in the first-come-first-serve manner.
The forwarding of the value of the entropy function $H$ in step~\ref{step:PPDistrForward} is also not difficult in CCDE: the source $\RZ{V_i}$ in a finite linear source model can be represented by a $|V_i|$-column matrix, which determines the value of $H(X)$ for all $X \subseteq V_i$.

%

\section{Conclusion}

This paper proposed a $\PAR$ algorithm that reduces the complexity of solving the minimum sum-rate problem and determining the PSP in communication for omniscience and other related problems by a factor of $|V|$.
We observed the existing $\CoordSatCapFus$ algorithm that determines the Dilworth truncation $\FuHat{\alpha}(V)$ in the minimum sum-rate estimate $\alpha$, which is segmented by the critical values and also characterizes the PSP $\Set{\Patp{j} \colon j \in \Set{1,\dotsc,p}}$. We proved that the objective function in a SFM problem in $\CoordSatCapFus$ exhibits strict strong map property so that the minimizer for all $\alpha$ is found by $O(1)$ calls of the PSFM algorithm that completes in $O(\SFM(|V|))$ time. Based on this fact, we proposed a $\PAR$ algorithm that obtains the PSP in $O(|V| \cdot \SFM(|V|))$ time.
We showed the distributed implementation of $\PAR$ by proposing the $\DistrPAR$ algorithm, which iteratively adapts the Dilworth truncation $\FuHat{\alpha}(V_i)$ of the subsystem $V_i$ for all $\alpha$ as $i$ increases. It converges to $\FuHat{\alpha}(V)$ finally where the first critical point $\alphap{1}$ and partition $\Patp{1}$ provide the solutions to the minimum sum-rate problem for both asymptotic and non-asymptotic models.
The PSP returned by the $\PAR$ or $\DistrPAR$ algorithm also provides the solutions to the secret key agreement, optimal network attack, information-theoretic and $\beta$-MAC clustering problems.

In addition to the brief discussion on the related problems in Section~\ref{sec:Relation}, it is worth studying how the $\PAR$ algorithm contributes to the recent developments in secret key agreement problem in \cite{Mukherjee2016IT,Chan2017ISIT} and the agglomerative approach for the information-theoretic clustering problem in \cite{AggloClust2018ISIT}.
The study also highlighted the importance of the PSFM algorithm. Given the fact that the PSFM algorithm is adapted from an existing SFM algorithm, it is worth discussing whether the minimum norm algorithm \cite{FujishigeMiniNorm}, which is the most practically efficient SFM algorithm, can also be adapted to a parametric one.

\appendices

\section{Properties of PSP in $\alpha$ and Decomposition Algorithm}
\label{app:AlphaAdapt}

For $f$ being a submodular function, e.g., the entropy $H$ or the cut $\kappa$ function. The solution to the minimization $\min_{\Pat \in \Pi(V)} \Fu{\lambda}[\Pat]$, where $\Fu{\lambda}[\Pat] = \sum_{C \in \Pat} \Fu{\lambda}(C)$ and $\Fu{\lambda}(C) = f(C) - \lambda$ , is segmented in $\lambda$ by critical points $\lambdap{j}$, or $\alphap{j} = H(V) - \lambdap{j}$, and $\Set{\Patp{j} \colon \Set{0,\dotsc,p}}$ as described in Section~\ref{subsec:PSP}. The $\lambdap{j}$'s and $\Patp{j}$ satisfy the following lemma.

\begin{lemma}[{\cite[Sections~2.2 and 3]{MinAveCost}\cite[Definition~3.8]{Narayanan1991PLP}}] \label{lemma:AlphaAdapt}
    For any two $\Patp{j}$ and $\Patp{j'}$ such that $j < j'$ (or $\Patp{j'} \prec \Patp{j}$), let
    $$ \lambda = \frac{ f[\Patp{j'}] - f[\Patp{j}] }{ |\Patp{j'}| - |\Patp{j}| } $$
    and $\alpha = f(V) - \lambda$. The followings hold.
    \begin{enumerate}[(a)]
      \item If $j + 1 = j'$, $\lambda = \lambdap{j'}$ and $\alpha = \alphap{j'}$;
      \item If $j + 1 < j'$, $\lambdap{j} \leq \lambda < \lambdap{j'}$ and $\alphap{j'} < \alpha \leq \alphap{j}$. \hfill \IEEEQED
    \end{enumerate}
\end{lemma}
Based on Lemma~\ref{lemma:AlphaAdapt}, the call $\DA(\Set{\Set{i} \colon i \in V},\Set{V})$ of the decomposition algorithm (DA) in Algorithm~\ref{algo:DA} returns all partitions in $\Set{\Patp{j} \colon j \in \Set{0,\dotsc,p}}$ of the PSP. The corresponding critical points $\alphap{j}$ or $\lambdap{j}$ can be determined by Lemma~\ref{lemma:AlphaAdapt}(a).
The MDA algorithm in \cite[Algorithm~1]{Ding2018IT} is a revised version of the DA algorithm for the purpose of determining only the first partition $\Patp{1}$, which determines the solution to the minimum sum-rate problem in CO.
Lemma~\ref{lemma:AlphaAdapt} also ensures the validity of $\StrMap$ algorithm in Algorithm~\ref{algo:StrMap}.

      \begin{algorithm} [t]
	       \label{algo:DA}
	       \small
	       \SetAlgoLined
	       \SetKwInOut{Input}{input}\SetKwInOut{Output}{output}
	       \SetKwFor{For}{for}{do}{endfor}
            \SetKwRepeat{Repeat}{repeat}{until}
            \SetKwIF{If}{ElseIf}{Else}{if}{then}{else if}{else}{endif}
	       \BlankLine
           \Input{$\Patp{j},\Patp{j'}$ in the PSP of $V$ such that $\Patp{j'} \prec \Patp{j}$.}
	       \Output{$\Set{\Patp{j},\Patp{j+1},\dotsc,\Patp{j'}}$.}
	       \BlankLine
            $\alpha \coloneqq H(V) - \frac{ f[\Patp{j'}] - f[\Patp{j}] }{ |\Patp{j'}| - |\Patp{j}| }$\;
            $(\rv_{\alpha,V},\Qat{\alpha}{V}) \coloneqq \CoordSatCapFus(\alpha,f,V,\Phi)$ where $\Phi$ is an arbitrarily chosen linear ordering of $V$\;
            \lIf{$\Qat{\alpha}{V} = \Patp{j'}$}{\Return $\Set{\Patp{j},\Patp{j'}}$}
            \lElse{ \Return $\DA(\Patp{j},\Qat{\alpha}{V}) \cup \DA(\Qat{\alpha}{V},\Patp{j'})$}
	       \caption{Decomposition Algorithm (DA) \cite[Algorithm~SPLIT]{MinAveCost} \cite[Algorithm~II]{Narayanan1991PLP}}
	   \end{algorithm}

\section{Proof of Lemma~\ref{lemma:EssProp}}
\label{app:lemma:EssProp}

The fact $\rv_{\alpha,V} \in P(\Fu{\alpha})$ holds throughout Algorithm~\ref{algo:CoordSatCapFus} is shown in \cite[Section~4.2]{MinAveCost} \cite[Lemma~19]{Ding2018IT} and the equality of two polyhedra $ P(\Fu{\alpha}) = P(\FuHat{\alpha})$ is proved in \cite[Theorem 25]{Fujishige2005}. (a) is the result in \cite[Theorem~8 and Lemma~9]{MinAveCost}.

We prove (b) and (c) as follows. All $C \in \Qat{\alpha}{V_i}$ are \emph{tight sets} \cite[Section~4.2]{MinAveCost}, i.e., $r_{\alpha}(C) = \Fu{\alpha}(C), \forall C \in \Qat{\alpha}{V_i}$.
In addition, for each $C \in \Qat{\alpha}{V_i}$, $r_\alpha(C) = \Fu{\alpha}(C) \leq \FuHat{\alpha}(C)$ since $\rv_{\alpha,V} \in P(\Fu{\alpha}) = P(\FuHat{\alpha})$. But, $\FuHat{\alpha}(C) \leq \Fu{\alpha}(C)$, too, based on the definition of Dilworth truncation~\eqref{eq:Dilworth}. So, $r_\alpha(C) = \Fu{\alpha}(C) = \FuHat{\alpha}(C)$ for all $C \in \Qat{\alpha}{V_i}$ and therefore $r_\alpha(\TX) = r_\alpha[\X] = \Fu{\alpha}[\X] = \FuHat{\alpha}[\X]$ for all $ \X \subseteq \Qat{\alpha}{V_i}$.
We also have $\X = \bigwedge \argmin_{\Pat \in \Pi(\TX)} \Fu{\alpha}[\Pat],\forall \X \subseteq \Qat{\alpha}{V_i}$ because, otherwise, either $\Qat{\alpha}{V_i} \notin \argmin_{\Pat \in \Pi(V_i)} \Fu{\alpha}[\Pat]$ or $\Qat{\alpha}{V_i}$ is not the finest minimizer. Therefore, (b) holds. (c) also holds because of the properties of the PSP in Section~\ref{subsec:PSP}. \hfill\IEEEQED

\section{Alternative to $\StrMap$ for $\DistrPAR$ Algorithm}
\label{app:PPAlt}

Similar to the $\StrMap$ algorithm, we derive the properties in Lemma~\ref{lemma:PPAlt} below and propose the $\StrMapDistPAR$ algorithm in Algorithm~\ref{algo:StrMapAlt} for determining the minimal minimizer $\U{\lambda}{V_i}$ of $\min\Set{ \FuU{\lambda_j}(\TX) \colon \Set{\phi_i} \in \X \subseteq \Qat{\lambda_j}{V_{i}}}$ in each iteration $i$ of the $\DistrPAR$ algorithm (Algorithm~\ref{algo:DistrParAlgo}).

\begin{lemma}\label{lemma:PPAlt}
    Consider the critical points $\Set{\lambda_j \colon j \in \Set{0,\dotsc,q}}$ and $\Set{\TUp{j} \colon j \in \Set{0,\dotsc,q}}$ that determine the minimal minimizer $\U{\lambda}{V_i}$ of $\min\Set{ \FuU{\lambda}(\TX) \colon \Set{\phi_i} \in \X \subseteq \Qat{\lambda}{V_i}}$ in Algorithm~\ref{algo:DistrParAlgo}.
    The followings hold:
    \begin{enumerate}[(a)]
        \item $\FuU{\lambda_j} (\TUp{j-1}) = \FuU{\lambda_j} (\TUp{j})$ for all $j \in \Set{1,\dotsc,q}$;
        \item For any $j,j' \in \Set{0,\dotsc,q}$ such that $j < j'$, let
            \begin{equation} \label{eq:Lambda}
                \lambda = \frac{ f[\ASet{\TUp{j}\setminus\TUp{j'}}{\Pat_d}] + f(\TUp{j'}) - f(\TUp{j}) }{ |\ASet{\TUp{j}\setminus\TUp{j'}}{\Pat_d}|},
            \end{equation}
            where $\Pat_d \preceq \Qat{\lambda_{j'}}{V_i}$.
            \begin{enumerate}[(i)]
                \item If $\Pat_d \prec \Qat{\lambda_{j'}}{V_i}$, let $\Patp{l}$ in the PSP of $V_{i-1}$ such that $\Pat_d = \Patp{l} \sqcup \Set{\Set{\phi_i}}$. Then, the corresponding critical value $\lambdap{l} > \lambda_{j'}$ and $\lambda_{j+1} \leq \lambda < \lambdap{l}$;
                \item If $\Pat_d = \Qat{\lambda_{j'}}{V_i}$, then $\lambda_{j+1} \leq \lambda < \lambda_{j'}$ for $j+1 < j'$ and $\lambda = \lambda_{j+1}$ for $j+1 = j'$.
            \end{enumerate}
    \end{enumerate}
\end{lemma}
\begin{IEEEproof}
    (a) is a result in \cite[Theorem~31]{Fujishige2009PP} of the strict strong map: for $\Up{j} = \bigcap \argmin\Set{ \FuU{\lambda_j}(\TX) \colon \Set{\phi_i} \in \X \subseteq \Qat{\lambda_j}{V_{i - 1}} \sqcup \Set{\Set{\phi_i}}}$ and $\Up{j-1} = \bigcup \argmin\Set{ \FuU{\lambda_j}(\TX) \colon \Set{\phi_i} \in \X \subseteq \Qat{\lambda_j}{V_{i - 1}} \sqcup \Set{\Set{\phi_i}}}$ for all $j \in \Set{1,\dotsc,q}$.

    By converting $\FuU{\lambda_j} (\TUp{j-1}) = \FuU{\lambda_j} (\TUp{j})$ to $r_{\lambda_j} (\TUp{j-1}\setminus\TUp{j}) = f(\TUp{j-1}) - f(\TUp{j})$ and \eqref{eq:Lambda} to $f(\TUp{j}) - f(\TUp{j'}) =  \Fu{\lambda}[\ASet{\TUp{j}\setminus\TUp{j'}}{\Pat_{d}}]$, we have $\sum_{m = j+1}^{j'} r_{\lambda_m} (\TUp{m-1}\setminus\TUp{m}) = \Fu{\lambda}[\ASet{\TUp{j}\setminus\TUp{j'}}{\Pat_{d}}]$ for $j < j'$.
    We prove (b) by contradiction. If $\Pat_d \prec \Qat{\lambda_{j'}}{V_i}$ and let $\Patp{l}$ be one of the partitions in the PSP of $V_i$, which characterize the segmented $\Qat{\lambda}{V_{i-1}}$ for all $\lambda$, such that $\Patp{l} \sqcup \Set{\Set{\phi_i}} = \Pat_d$, then we must have $\lambdap{l} > \lambda_{j'}$.
    If $\lambda < \lambda_{j+1}$, we have
    $ \sum_{m = j+1}^{j'} r_{\lambda_m} (\TUp{m-1}\setminus\TUp{m}) > \Fu{\lambda_{j+1}}[\ASet{\TUp{j}\setminus\TUp{j'}}{\Pat_{d}}] \geq \sum_{m = j+1}^{j'} \Fu{\lambda_{m}}[\ASet{\TUp{m-1}\setminus\TUp{m}}{\Pat_d}]$
    contradicting $\rv_{\lambda,V} \in P(f_\lambda), \forall \lambda \in H(V) - \alpha$ in Lemma~\ref{lemma:EssProp}; If $\lambda \geq \lambdap{l}$, we have $ \sum_{m = j+1}^{j'} r_{\lambda_m} (\TUp{m-1}\setminus\TUp{m}) \leq \Fu{\lambdap{l}}[\ASet{\TUp{j}\setminus\TUp{j'}}{\Pat_{d}}] = \FuHat{\lambdap{l}} (\TUp{j}\setminus\TUp{j'}) < \FuHat{\lambda_{j'}} (\TUp{j}\setminus\TUp{j'}) \leq \sum_{m = j+1}^{j'} \FuHat{\lambda_{j'}}(\TUp{m-1}\setminus\TUp{m}) \leq \sum_{m = j+1}^{j'} \FuHat{\lambda_{m}}(\TUp{m-1}\setminus\TUp{m})$ contradicting $r_\lambda(\TX) = \FuHat{\lambda} (\TX)$ for all $\X \subseteq \Qat{\lambda}{V_i}$ and $\lambda = H(V) - \alpha$ in Lemma~\ref{lemma:EssProp}(b). So, we must have $\lambda_{j+1} \leq \lambda < \lambdap{l}$ and (b)-(i) holds.

    For $\Pat_d = \Qat{\lambda_{j'}}{V_i}$, consider the case when $ j + 1 < j'$. Assume that $\lambda < \lambda_{j+1}$. Then, we have $ \sum_{m = j+1}^{j'} r_{\lambda_m} (\TUp{m-1}\setminus\TUp{m}) > \Fu{\lambda_{j+1}}[\ASet{\TUp{j}\setminus\TUp{j'}}{\Pat_{\lambda_{j'}}}] > \sum_{m = j+1}^{j'} \Fu{\lambda_{m}}[\ASet{\TUp{m-1}\setminus\TUp{m}}{\Pat_{\lambda_{j'}}}]$ contradicting $\rv_{\lambda,V} \in P(f_\lambda), \forall \lambda = H(V) - \alpha$ in Lemma~\ref{lemma:EssProp}. Assume that $\lambda \geq \lambda_{j'}$. Then, we have $ \sum_{m = j+1}^{j'} r_{\lambda_m} (\TUp{m-1}\setminus\TUp{m}) \leq \Fu{\lambda_{j'}}[\ASet{\TUp{j}\setminus\TUp{j'}}{\Pat_{\lambda_{j'}}}] = \FuHat{\lambda_{j'}} (\TUp{j}\setminus\TUp{j'}) \leq \sum_{m = j+1}^{j'} \FuHat{\lambda_{j'}}(\TUp{m-1}\setminus\TUp{m}) < \sum_{m = j+1}^{j'} \FuHat{\lambda_{m}}(\TUp{m-1}\setminus\TUp{m}) $ contradicting $r_\lambda(\TX) = \FuHat{\lambda} (\TX)$ for all $\X \subseteq  \Pat_\lambda$ and $\lambda$ in Lemma~\ref{lemma:EssProp}(b). Therefore, we must have $\lambda_{j+1} \leq \lambda < \lambda_{j'}$.
    Consider the case when $ j+1 = j'$. Assume that $\lambda < \lambda_{j'}$. Then, we have $ r_{\lambda_{j'}} (\TUp{j}\setminus\TUp{j'}) > \Fu{\lambda_{j'}}[\ASet{\TUp{j}\setminus\TUp{j'}}{\Pat_{\lambda_{j'}}}] $ contradicting $\rv_{\lambda_{j'},V} \in P(f_{\lambda_{j'}})$ in Lemma~\ref{lemma:EssProp}. Assume $\lambda > \lambda_{j'}$. Then, we have $ r_{\lambda_{j'}} (\TUp{j}\setminus\TUp{j'}) < \Fu{\lambda_{j'}}[\ASet{\TUp{j}\setminus\TUp{j'}}{\Pat_{\lambda_{j'}}}] = \FuHat{\lambda_{j'}}(\TUp{j}\setminus\TUp{j'}) $ contradicting $r_{\lambda_{j'}}(\TX) = \FuHat{\lambda_{j'}} (\TX), \forall \X \subseteq \Pat_{\lambda_{j'}}$ in Lemma~\ref{lemma:EssProp}(b). Therefore, we must have $\lambda = \lambda_{j'}$. (b)-(ii) holds.
\end{IEEEproof}

The call $\StrMapDistPAR(V_i,\Set{\phi_i},\Set{\Set{m} \colon m \in V_i})$ returns $\Set{\TUp{j} \colon j \in \Set{0,\dotsc,q}}$ that segments the minimal minimizer $\U{\lambda}{V_i}$ of $\min \Set{ \FuU{\lambda}(\TX) \colon \Set{i} \in \X \subseteq \Qat{\lambda}{V_i}}$ for all $\lambda$ in each iteration $i$ of the $\DistrPAR$ algorithm. The corresponding critical points $\lambda_{j}$'s can also be determined by Lemma~\ref{lemma:PPAlt}(a).
By replacing $\lambda$ with $f(V) - \alpha$, the $\StrMapDistPAR$ algorithm can be applied to determine $\U{\alpha}{V_i}$ for all $\alpha$ in step~\ref{step:PP} of $\PAR$ algorithm. The $\StrMapDistPAR$ returns the same results as $\StrMap$ and can also be implemented by the PSFM algorithms in \cite{Fleischer2003PSFM,Nagano2007PSFM,IwataPSFM1997}.

       \begin{algorithm} [t]
	       \caption{$\StrMapDistPAR(\TUp{j}, \TUp{j'},\Pat_d)$: Find $\Set{\TUp{j} \colon j \in \Set{0,\dotsc,q}}$ in step~\ref{step:PPDistrPar} of the $\DistrPAR$ algorithm (Algorithm~\ref{algo:DistrParAlgo})}\label{algo:StrMapAlt}
	       \small
	       \SetAlgoLined
	       \SetKwInOut{Input}{input}\SetKwInOut{Output}{output}
	       \SetKwFor{For}{for}{do}{endfor}
           \SetKwRepeat{Repeat}{repeat}{until}
           \SetKwIF{If}{ElseIf}{Else}{if}{then}{else if}{else}{endif}
	       \BlankLine
           \Input{$\TUp{j}, \TUp{j'}$ such that $\TUp{j} \supseteq \TUp{j'}$ and $\Pat_d$.}
	       \Output{$\Set{\TUp{j}, \TUp{j+1}, \dotsc, \TUp{j'}}$.}
	       \BlankLine
           \lIf{$\TUp{j} = \TUp{j'}$}{\Return $\Set{\TUp{j}}$}
           \Else{
                $\lambda \coloneqq \frac{  f[\ASet{\TUp{j} \setminus \TUp{j'}}{\Pat_d}] + f(\TUp{j'}) - f(\TUp{j})}{ |\ASet{\TUp{j} \setminus \TUp{j'}}{\Pat_d}|}$\;
                $\U{\lambda}{V_i} \coloneqq \bigcap \argmin \Set{ \FuU{\lambda}(\TX) \colon \Set{\phi_i} \in \X \subseteq \Qat{\lambda}{V_i}}$\;
                \lIf{$\TUp{j'} = \TU{\lambda}{V_i}$ and $\FuU{\lambda}(\TUp{j}) = \FuU{\lambda}(\TU{\lambda}{V_i})$}{\Return $\Set{\TUp{j}, \TUp{j'}}$}
                \lElse{\Return $\StrMapDistPAR(\TUp{j},\TU{\lambda}{V_i},\Qat{\lambda}{V_i}) \cup \StrMapDistPAR(\TU{\lambda}{V_i},\TUp{j'},\Pat_d) $}
                }
	   \end{algorithm}

      \begin{algorithm} [t]
	       \caption{$\StrMapDistPAR(d_{\TUp{j}},S_u,d_{S_d},S_d,\Pat_d)$ by the PSFM \cite{Fleischer2003PSFM}} \label{algo:PSFM}
	       \small
	       \SetAlgoLined
	       \SetKwInOut{Input}{input}\SetKwInOut{Output}{output}
	       \SetKwFor{For}{for}{do}{endfor}
           \SetKwRepeat{Repeat}{repeat}{until}
           \SetKwIF{If}{ElseIf}{Else}{if}{then}{else if}{else}{endif}
	       \BlankLine
           \Input{$\TUp{j}, \TUp{j'}$ such that $\TUp{j} \supseteq \TUp{j'}$, labelling $d_{\TUp{j}}$ and $d_{\TUp{j'}}$ associated with $\TUp{j}$ and $\TUp{j‘}$, respectively, and $\Pat_d$.}
	       \Output{$\Set{\TUp{j}, \TUp{j+1}, \dotsc, \TUp{j'}}$.}
	       \BlankLine
           \lIf{$\TUp{j} = \TUp{j'}$}{\Return $\Set{\TUp{j}}$}
           \Else{
                $\lambda \coloneqq \frac{  f[\ASet{\TUp{j} \setminus \TUp{j'}}{\Pat_d}] + f(\TUp{j'}) - f(\TUp{j})}{ |\ASet{\TUp{j} \setminus \TUp{j'}}{\Pat_d}|}$\;
                Run the push-relabel SFM in \cite[Fig.~1]{Fleischer2003PSFM} to determine $\U{\lambda}{V_i} \coloneqq \bigcap \argmin \Set{ \FuU{\lambda}(\TX) \colon \Set{\phi_i} \in \X \subseteq \Qat{\lambda}{V_i}}$, and the resulting labels labelling $d_{\U{\lambda}{V_i}}$; At the same time, run the reverse-push-relabel SFM of \cite[Fig.~1]{Fleischer2003PSFM} to determine $\U{\lambda}{V_i}$ and the resulting labelling $d_{\U{\lambda}{V_i}}^R$\;
                \lIf{$\TUp{j'} = \TU{\lambda}{V_i}$ and $\FuU{\lambda}(\TUp{j}) = \FuU{\lambda}(\TU{\lambda}{V_i})$}{\Return $\Set{\TUp{j}, \TUp{j'}}$}
                \lElse{
                     If the push-relabel SFM terminates first and $|\TU{\lambda}{V_i}| > \frac{|V_i|}{2}$, then \Return $\StrMapDistPAR(d_{\TUp{j}},\TUp{j},d_0,\TU{\lambda}{V_i},\HPat{\lambda}) \cup \StrMapDistPAR(d_{\SCal},\TU{\lambda}{V_i},d_{\TUp{j'}},\TUp{j'},\Pat_d)$, where $d_0$ is all zero labelling; otherwise wait until the reverse-push-relabel SFM completes and \Return $\StrMapDistPAR(d_{\TUp{j}},\TUp{j},d_{\SCal}^R,\TU{\lambda}{V_i},\HPat{\lambda}) \cup \StrMapDistPAR(d_0,\TU{\lambda}{V_i},d_{\TUp{j'}},\TUp{j'},\Pat_d)$.
                     Or, if the reverse-push-relabel SFM terminates first and $|\TU{\lambda}{V_i}| \leq \frac{|V_i|}{2}$, then \Return $\StrMapDistPAR(d_{\TUp{j}},\TUp{j},d_{\SCal}^R,\TU{\lambda}{V_i},\HPat{\lambda}) \cup \StrMapDistPAR(d_0,\TU{\lambda}{V_i},d_{\TUp{j'}},\TUp{j'},\Pat_d)$; otherwise until the push-relabel SFM completes and \Return $\StrMapDistPAR(d_{\TUp{j}},\TUp{j},d_0,\TU{\lambda}{V_i},\HPat{\lambda}) \cup \StrMapDistPAR(d_{\SCal},\TU{\lambda}{V_i},d_{\TUp{j'}},\TUp{j'},\Pat_d)$
               }
          }
	   \end{algorithm}

\section{$\StrMap$ by PSFM}
\label{app:PSFM}

In \cite{ParMaxFlow1989}, the push-relabel $\MaxFlow$ algorithm in \cite{MaxFlow1988} was extended to a parameterized one based on the fact: if the capacities of edges from the source node and to the sink node are monotonically changing with a real-valued parameter $\alpha$, the max-flows/min-cuts for a finite number of monotonic values of $\alpha$ can be determined in the same asymptotic time as the push-relabel $\MaxFlow$ algorithm.
The same technique was further applied to extend the SFM algorithms to the PSFM ones in \cite{Fleischer2003PSFM,Nagano2007PSFM,IwataPSFM1997}. But, all PSFMs in \cite{Fleischer2003PSFM,Nagano2007PSFM,IwataPSFM1997} requires a finite number of monotonic values of $\alpha$ as the inputs. For solving the problem  $\min\Set{ \FuU{\lambda}(\TX) \colon \Set{\phi_i} \in \X \subseteq \Qat{\lambda}{V_i}}$ where the critical values of $\alpha$ are not known in advance, the $\StrMap$ algorithm can be implemented in the same way as the Slicing algorithm in \cite[Section~4.2]{Fleischer2003PSFM}.

We show in Algorithm~\ref{algo:PSFM} how to implement the $\StrMapDistPAR$ algorithm by the Slicing algorithm in \cite[Section~4.2]{Fleischer2003PSFM}. 
The PSFM proposed in \cite[Section~4.2]{Fleischer2003PSFM} is an extension of the Schrijver's SFM algorithm \cite{Schrijver2003} for the strong map sequence. In \cite[Fig.~1]{Fleischer2003PSFM}, the Schrijver's SFM algorithm \cite{Schrijver2003} is nested in a push-relabel framework so that, for all $\lambda < \lambda'$, the distance labels for determining the minimal minimizer $\U{\lambda}{V_i}$ of the problem $\min\Set{ \FuU{\lambda}(\TX) \colon \Set{\phi_i} \in \X \subseteq \Qat{\lambda}{V_i}}$ are still valid and can be reused to determine the minimal minimizer $\U{\lambda'}{V_i}$ of $\min\Set{ \FuU{\lambda'}(\TX) \colon \Set{\phi_i} \in \X \subseteq \Qat{\lambda'}{V_i}}$; 
On the other hand, the minimal minimizer $\U{\lambda'}{V_i}$ determined by the reverse-push-relabel SFM \cite[Fig.~1]{Fleischer2003PSFM} are valid and can be reused to determine $\U{\lambda}{V_i}$.
Thus, all $\TUp{j}$'s in the PP  can be determine by either the push-relabel SFM \cite[Section~4.2]{Fleischer2003PSFM} in ascending order of $\lambda_j$ or the reverse-push-relabel SFM of \cite[Section~4.2]{Fleischer2003PSFM} in descending $\lambda_j$ in the same asymptotic time as the original Schrijver's SFM algorithm \cite[Section~2.2.1]{Fleischer2003PSFM}.
In the case when the critical points $\lambda_j$'s are unknown, the Slice algorithm in \cite[Section~4.2]{Fleischer2003PSFM} suggests simultaneously implement the push-relabel SFM \cite[Fig.~1]{Fleischer2003PSFM} and its reverse in the recursion of the $\StrMapDistPAR$ algorithm as in Algorithm~\ref{algo:PSFM}.
In the initial call $\StrMapDistPAR(d_{V_i},V_i,d_{\Set{i}},\Set{i},\Set{\Set{m} \colon m \in V_i})$, $d_{V_i}$ and $d_{\Set{i}}$ can be both set to zero labelling, or, one can refer to \cite[Section~4.2]{Fleischer2003PSFM} as to how to properly select the value of $\lambda$ to set $d_{V_i}$ and $d_{\Set{i}}$.

\section{Kolmogorov's Algorithm}
\label{app:Kolmogorov}

For determining the PSP of a graph, Kolmogorov proposed Algorithm~\ref{algo:Komolgorov} in \cite[Fig.~3]{Kolmogorov2010} with the parametric $\MaxFlow$ \cite{ParMaxFlow1989} being the subroutine, the contribution of which is similar to the $\PAR$ algorithm: it reduces the previous complexity $O(|V|^2 \cdot \MaxFlow(|V|))$ for determining the network strength and the maximum number of edge-disjoint spanning trees \cite{Cunningham1985NetStrength,IBMRep2011} to $O(|V| \cdot \MaxFlow(|V|))$.
Algorithm~\ref{algo:Komolgorov} is base on the SFM problem $ \min \Set{ \FuUK{\lambda}(X) \colon \phi_i \in X \subseteq V } $ for\footnote{The set function $\FuUK{\lambda}$ in \eqref{eq:FusObjKomo} differs from $\FuU{\lambda}$ in Algorithm~\ref{algo:DistrParAlgo} in that $\FuUK{\lambda}$ does not merge the dimensions $i \in C$ in one subset $C \in \Qat{\lambda}{V}$.}
\begin{equation}\label{eq:FusObjKomo}
    \FuUK{\lambda}(X) \coloneqq \Fu{\lambda} (X) - r_\lambda (X), \qquad \forall X \subseteq V,
\end{equation}
where the rate vector $\rv_{\lambda}$ is updated in steps~\ref{step:UpdateStartKomo} and \ref{step:UpdateEndKomo} in each iteration to maintain the monotonicity: $r_{\lambda,i} \leq r_{\lambda',i}$ for all $i \in V$ and $\lambda < \lambda'$.
It is show in \cite[Lemmas~4 and 5]{Kolmogorov2010} that the minimizer of $ \min \Set{ \FuUK{\lambda}(X) \colon \phi_i \in X \subseteq V } $ forms a `nesting' set sequence in $\lambda$, which, for $f$ being the cut function, can be determined by only one call of the parametric $\MaxFlow$ algorithm in \cite{ParMaxFlow1989}.
We show in Theorem~\ref{theo:StrongMapK} and Lemma~\ref{lemma:PPpropKomo} below that this `nesting' property is also due to the strong map property, which is conditioned on the monotonicity of $\rv_{\lambda}$.

\begin{theorem} \label{theo:StrongMapK}
    In each iteration $i$ of Algorithm~\ref{algo:Komolgorov}, $\FuUK{\lambda}$ forms a \textbf{non-strict strong map} sequence in $\lambda$, i.e., $\FuUK{\lambda} \leftarrow \FuUK{\lambda'}$ for all $\lambda$ and $\lambda'$ such that $\lambda < \lambda'$.
\end{theorem}
\begin{IEEEproof}
    For any $X, Y \subseteq 2^V$ such that $X \subseteq Y$ and $\phi_i \notin Y \setminus X$, we have $\FuUK{\lambda}(Y) - \FuUK{\lambda}(X) - \FuUK{\lambda'}(Y) + \FuUK{\lambda'}(X) = r_{\lambda'}(Y \setminus X) - r_{\lambda}(Y \setminus X) \leq 0$.
    But, this inequality does not hold strictly for all $X \subsetneq Y$ such that $\phi_i \notin Y \setminus X$ since $\rv_{\lambda,V}$ is only nonincreasing, instead of strictly increasing, in $\lambda$. Based on Definition~\ref{def:StrMap}, theorem holds.
\end{IEEEproof}

\begin{lemma}{\cite[Theorems 26 to 28]{Fujishige2009PP}} \label{lemma:PPpropKomo}
    In each iteration $i$ of the Algorithm~\ref{algo:Komolgorov}, the minimal minimizer $U_{\lambda,V} = \bigcap \argmin \Set{ \FuUK{\lambda}(X) \colon \phi_i \in X \subseteq V }$ satisfies $U_{\lambda,V} \supseteq U_{\lambda',V}$ for all $\lambda < \lambda'$. $U_{\lambda,V}$ for all $\lambda$ is fully characterized by $q' < |V| - 1$ critical points
    $$0 = \lambda_0 < \lambda_1 < \dotsc < \lambda_{q'} < \lambda_{q'+1} = +\infty$$
    and the corresponding minimal minimizer $\UKomop{j} = U_{\lambda_j,V}$ forms a set chain
    $$ V = \UKomop{0} \supsetneq \UKomop{1} \supsetneq  \dotsc \supsetneq \UKomop{q'} = \Set{\phi_i}$$
    such that $U_{\lambda,V} = \UKomop{j}$ for all $\lambda \in [\lambda_j, \lambda_{j+1})$ and $j \in \Set{0,\dotsc,q'}$.   \hfill\IEEEQED
\end{lemma}

      \begin{algorithm} [t]
	       \caption{Komolgorov's Algorithm \cite[Fig.~3]{Kolmogorov2010}} \label{algo:Komolgorov}
	       \SetAlgoLined
	       \SetKwInOut{Input}{input}\SetKwInOut{Output}{output}
	       \SetKwFor{For}{for}{do}{endfor}
            \SetKwRepeat{Repeat}{repeat}{until}
            \SetKwIF{If}{ElseIf}{Else}{if}{then}{else if}{else}{endif}
	       \BlankLine
           \Input{$f$, $V$ and $\Phi$.}
	       \Output{$\Qat{\lambda}{V} = \bigwedge \argmin_{\Pat\in \Pi(V)} \Fu{\alpha}[\Pat]$ and $\rv_\lambda \in B(\FuHat{\lambda})$ for all $\lambda$.}
	       \BlankLine
                Initiate $ \rv_{\lambda,V} \coloneqq (-\lambda,\dotsc,-\lambda)$ and $\Qat{\lambda}{V}\coloneqq \Set{\Set{i}: i \in V}$ for all $\lambda$\;
                \For{$i=1$ \emph{\KwTo} $|V|$}{
                    For function $\FuUK{\lambda}(X) \coloneqq \Fu{\lambda} (X) - r_\lambda (X)$, obtain the critical points $\Set{\lambda_j \colon j \in \Set{0,\dotsc,q'}}$ and $\Set{\UKomop{j} \colon j \in \Set{0,\dotsc,q'}}$ that determine the minimal minimizer $U_{\lambda,V}$ of $\min \Set{ \FuUK{\lambda}(X) \colon \phi_i \in X \subseteq V }$\label{step:PPKomo}\;
                    Update $r_{\lambda,\phi_i} \coloneqq \Fu{\lambda}(U^{(j)})$ for all $\lambda \in [\lambda_j, \lambda_{j+1})$\label{step:UpdateStartKomo}\;
                    \lForEach{$m \in V$ such that $m \neq \phi_i$}{
                        find $j^* \in \Set{0,\dotsc,q}$ such that $m \in U_{j^* - 1} \setminus U_{j^*}$ and update $r_{\lambda,m} \coloneqq  \min \Set{ r_{\lambda,m}, r_{\lambda_{j^*},m} }$ for all $\lambda$ \label{step:UpdateEndKomo}
                    }
                    \lFor{$j=0$ \emph{\KwTo} $q$}{
                        $$ \U_{\lambda,V} \coloneqq \Set{C \in \Qat{\lambda}{V} \colon C \cap \UKomop{j} \neq \emptyset }, $$
                        $$ \Qat{\lambda}{V} \coloneqq (\Qat{\lambda}{V} \setminus \U_{\lambda,V}) \cup \Set{\TU_{\lambda,V}} $$
                        for all $\lambda \in [\lambda_j, \lambda_{j+1})$\label{step:UpdatePartsKomo}
                    }
                }
                \Return $\Qat{\lambda}{V}$ and $\rv_\lambda$ for all $\lambda$\;
	   \end{algorithm}

We derive the properties based on the strong map property in Theorem~\ref{theo:StrongMapK} below and show that Algorithm~\ref{algo:Komolgorov} can determine the PSP for the general submodular function $f$.

\begin{lemma}\label{lemma:PPKomo}
    In each iteration $i$ of Algorithm~\ref{algo:Komolgorov}, consider the critical points $\Set{\lambda_j \colon j \in \Set{1,\dotsc,q'}}$ and $\Set{\UKomop{j} \colon j \in \Set{1,\dotsc,q'}}$ that characterize the minimal minimizer $U_{\lambda,V}$ of $\min \Set{ \FuUK{\lambda}(X) \colon \phi_i \in X \subseteq V }$. For any two $j, j' \in \Set{0,\dotsc,q'}$ such that $j < j'$, let the value of $\lambda$ holds
    \begin{equation} \label{eq:LambdaK}
         r_{\lambda} (\UKomop{j} \setminus \UKomop{j'}) = f(\UKomop{j}) - f(\UKomop{j'}).
    \end{equation}
    The followings hold:
    \begin{enumerate}[(a)]
        \item If $j + 1 = j'$, $\lambda = \lambda_{j'}$;
        \item If $j + 1 <  j'$, $\lambda_{j+1} \leq \lambda \leq \lambda_{j'}$.
    \end{enumerate}
\end{lemma}
\begin{IEEEproof}
    For the monotonicity $\rv_{\lambda,V} \geq \rv_{\lambda,V}$ for all $\lambda < \lambda'$, we have $\lambda < \lambda'$ if $\rv_{\lambda,V} > \rv_{\lambda',V}$ and $\lambda \leq \lambda'$ if $\rv_{\lambda,V} \geq \rv_{\lambda',V}$.\footnote{We denote $\rv_{\lambda} \geq \rv_{\lambda}$ if $r_{\lambda,i} \geq r_{\lambda',i}$ for all $i \in V$ and $\rv_{\lambda,V} > \rv_{\lambda',V}$ if $r_{\lambda,i} \geq r_{\lambda,i}$ for all $i \in V$ and at least one inequality holds strictly.}
    Based on Lemma~\ref{lemma:PPpropKomo}, for $j < j'$ and a sufficiently small $\epsilon > 0$, we have $U_{\lambda_{j+1} -\epsilon,V} = \UKomop{j}$. Since $U_{\lambda,V}$ is the minimal minimizer of $\min \Set{ \FuUK{\lambda}(X) \colon \phi_i \in X \subseteq V }$ for all $\lambda$ and $\UKomop{j} \supsetneq \UKomop{j'}$, we have
    $\FuUK{\lambda_{j+1}-\epsilon}(\UKomop{j}) < \FuUK{\lambda_{j+1}-\epsilon}(\UKomop{j'})$ and $\FuUK{\lambda_{j'}}(\UKomop{j}) \geq \FuUK{\lambda_{j'}}(\UKomop{j'})$ such that
    $r_{\lambda_{j+1}-\epsilon} (\UKomop{j} \setminus \UKomop{j'}) > f(\UKomop{j}) - f(\UKomop{j'})$ and $r_{\lambda_{j'}} (\UKomop{j} \setminus \UKomop{j'}) \leq f(\UKomop{j}) - f(\UKomop{j'})$, respectively.

    For the value of $\lambda$ that satisfies \eqref{eq:LambdaK}, we have
    $$r_{\lambda_{j'}} (\UKomop{j} \setminus \UKomop{j'}) \leq r_{\lambda} (\UKomop{j} \setminus \UKomop{j'}) < r_{\lambda_{j+1}-\epsilon} (\UKomop{j} \setminus \UKomop{j'})$$
    Thus, $\lambda_{j+1} - \epsilon < \lambda \leq \lambda_{j'}$ for any sufficiently small $\epsilon$, which, due to the continuity of $\FuHat{\lambda}(V)$ in $\lambda$, is equivalent to $\lambda_{j+1} \leq \lambda \leq \lambda_{j'}$ and reduces to $\lambda = \lambda_{j'}$ in the case when $j+1 = j'$.
\end{IEEEproof}

Based on Lemma~\ref{lemma:PPKomo}, the call $\StrMapKomo(V,\Set{\phi_i})$ of Algorithm~\ref{algo:StrMapKomo} returns all $\UKomop{j}$'s that segments the minimal minimizer $U_{\lambda,V}$ of $\min \Set{ \FuUK{\lambda}(X) \colon \phi_i \in X \subseteq V }$ in each iteration $i$ of Algorithm~\ref{algo:Komolgorov}. The corresponding critical points $\lambda_{j}$'s can be obtained by Lemma~\ref{lemma:PPKomo}(a). Again, the $\StrMapKomo$ algorithm can be implemented by the PSFM algorithms \cite{Fleischer2003PSFM,Nagano2007PSFM,IwataPSFM1997}.
It should be noted that in Algorithm~\ref{algo:Komolgorov}, we need to check whether $U_{\lambda - \epsilon}$, the minimal minimizer of the problem $\min \Set{ \FuUK{\lambda-\epsilon}(X) \colon \phi_i \in X \subseteq V }$, equals to $\UKomop{j}$ for a small $\epsilon > 0$ before terminating the recursion.
The reason is to avoid missing the subsets $\UKomop{j''}$ such that $\UKomop{j} \supsetneq \UKomop{j''} \supsetneq \UKomop{j'}$ with the critical point $\lambda_{j''} \in (\lambda_j,\lambda_{j'})$.\footnote{In the case when $j + 1< j'$, we could have $\lambda = \lambda_{j'}$ as in Lemma~\ref{lemma:PPKomo}(b) with $U_{\lambda} = \UKomop{j'}$, where, if the recursion is terminated, the critical values $\lambda_{j''}$ will not be searched.}
Therefore, the value of $\epsilon$ should be chosen sufficiently small for the validity of Algorithm~\ref{algo:Komolgorov}. This is because the strong map property of $\FuUK{\lambda}$ is non-strict.

      \begin{algorithm} [t]
	       \caption{$\StrMapKomo(\UKomop{j},\UKomop{j'})$: Find $\Set{\UKomop{j} \colon j \in \Set{1,\dotsc,q'}}$ in step~\ref{step:PPKomo} of the Kolmogorov's algorithm (Algorithm~\ref{algo:Komolgorov})}\label{algo:StrMapKomo}
	       \SetAlgoLined
	       \SetKwInOut{Input}{input}\SetKwInOut{Output}{output}
	       \SetKwFor{For}{for}{do}{endfor}
           \SetKwRepeat{Repeat}{repeat}{until}
           \SetKwIF{If}{ElseIf}{Else}{if}{then}{else if}{else}{endif}
	       \BlankLine
           \Input{$\UKomop{j}, \UKomop{j'}$ such that $\UKomop{j} \supsetneq \UKomop{j'}$.}
	       \Output{$\Set{\UKomop{j}, \UKomop{j+1}, \dotsc, \UKomop{j'}}$.}
	       \BlankLine
                Determine $\lambda$ such that $r_{\lambda} (\UKomop{j} \setminus \UKomop{j'}) = f(\UKomop{j}) - f(\UKomop{j'})$\;
                $U_{\lambda,V} \coloneqq \bigcap \argmin \Set{ \FuUK{\lambda}(X) \colon \phi_i \in X \subseteq V }$\;
                \uIf{$U_{\lambda,V} = \UKomop{j'}$}{
                    Obtain $U_{\lambda - \epsilon,V} \coloneqq \bigcap \argmin \Set{ \FuUK{\lambda-\epsilon}(X) \colon \phi_i \in X \subseteq V }$ for a small $\epsilon > 0$\;
                    \lIf{$U_{\lambda - \epsilon,V} = \UKomop{j}$}{\Return $\Set{U^{(j)},U^{(j')}}$}
                    \lElse{\Return $\StrMapKomo(\UKomop{j},U_{\lambda - \epsilon,V}) \cup \StrMapKomo(U_{\lambda - \epsilon,V},\UKomop{j'}) $}
                    }
	   \end{algorithm}

\bibliographystyle{IEEEtran}
\bibliography{COSOBIB}

\end{document}